\documentclass[12pt]{article}
\usepackage{amssymb,amsmath}
\usepackage{bm} 
\usepackage{booktabs} 
\usepackage{array}
\usepackage{latexsym}
\usepackage{graphicx}
\usepackage{rotating}
\usepackage{color}
\usepackage{slashed}
\usepackage{datetime}
\usepackage[nosort]{cite}
\usepackage{verbatim}
\usepackage{enumerate}
\usepackage{chngpage} 
\allowdisplaybreaks
\usepackage{psfrag}
\usepackage{mathtools}
\usepackage{mciteplus}
\usepackage{dsfont}
\usepackage{multirow} 
\usepackage{subcaption}

\usepackage[colorlinks=true,      linkcolor=blue,      urlcolor=blue,      
            filecolor=blue,      citecolor=blue,       pdfstartview=FitH,     
						pdfpagemode=UseNone,      bookmarksopen=true]{hyperref}  
\usepackage[all]{hypcap}     

\def\be{\begin{equation}}
\def\ee{\end{equation}}
\def\bal{\begin{align}}
\def\eal{\end{align}}

\def\({\left(}
\def\){\right)}
\def\[{\left[}
\def\]{\right]}

\definecolor{cardinal}{rgb}{0.6,0,0}
\definecolor{darkgreen}{rgb}{0,0.4,0}
\definecolor{golden}{rgb}{0.92, 0.7, 0}
\definecolor{midnight}{rgb}{0, 0, 0.5}
\definecolor{darkblue}{rgb}{0, 0, 0.7}
\definecolor{purple}{rgb}{0.5, 0, 0.5}

\newcommand{\D}[2][\phantom{}]{\ensuremath{\text{D}#2_{#1}}}
\newcommand{\F}[1][\phantom{}]{\ensuremath{\text{F1}_{#1}}}
\newcommand{\W}[1][\phantom{}]{\ensuremath{\text{P}_{#1}}}
\newcommand{\NS}[1][\phantom{}]{\ensuremath{\text{NS5}_{#1}}}
\newcommand{\KKM}[1][\phantom{}]{\ensuremath{\text{KKM}_{#1}}}
\newcommand{\M}[2][\phantom{}]{\ensuremath{\text{M}#2_{#1}}}

\usepackage{epsf}
\usepackage{cite}
\usepackage{fancyhdr}
\usepackage{bm}
\usepackage{enumerate} 

\usepackage{empheq} \empheqset{box=\fbox}
\usepackage[usenames,dvipsnames]{xcolor}

\numberwithin{equation}{section}  
\allowdisplaybreaks

\usepackage{graphicx}
\usepackage{tikz}
\usetikzlibrary{decorations.markings}
\usetikzlibrary{shapes,shapes.geometric, arrows, positioning}
\usetikzlibrary{fadings}
\usetikzlibrary{external}
\usetikzlibrary{tikzmark}
\tikzset{->-/.style={decoration={
			markings,
			mark=at position #1 with {\arrow{stealth}}},postaction={decorate}}}
\tikzset{%
            base/.style = {rectangle, rounded corners, draw=black,
                           minimum width=2cm, minimum height=.8cm,
                           text centered},
             IIA/.style = {base, fill=red!20},
             IIB/.style = {base, fill=yellow!20},
            IIAB/.style = {base, fill=orange!20},
        TDuality/.style = {thick,->,>=stealth},
        SDuality/.style = {gray!70,thin,<->,>=stealth},
}
\usepackage{adjustbox}
\usepackage{pgfplots}
\pgfplotsset{compat=1.11}
\usepgfplotslibrary{fillbetween}
\usetikzlibrary{intersections}
\usetikzlibrary{patterns}

\pgfdeclarelayer{bg}
\pgfsetlayers{bg,main}

\begin{document}


\begin{flushright}
MPP-2023-291\\

\end{flushright}

\vspace{3mm}

\begin{center}

{\huge {\bf The 1/4-BPS building blocks of brane interactions}}

\vspace{14mm}

{\large
\textsc{Ben Eckardt $^{a,b}$, Yixuan Li $^{a}$}}
\vspace{12mm}

\textit{$^{a}$ Max-Planck-Institut f\"ur Physik (Werner-Heisenberg-Institut), \\
	Boltzmannstraße 8, 85748 Garching, Germany  \\}  
\medskip
\medskip
\textit{$^{b}$ Ludwig-Maximilians-Universit{\"a}t M\"unchen, Fakult{\"a}t f{\"u}r Physik,\\ 
Theresienstr.~37, 80333 M\"unchen, Germany  \\}  
\medskip

\vspace{4mm} 
%

{\footnotesize\upshape\ttfamily eckardt, yixuan @ mpp.mpg.de} 
\vspace{13mm}
 

\end{center}

\begin{adjustwidth}{10mm}{10mm} 

\begin{abstract}
\vspace{1mm}
\noindent

We study, from the perspective of supersymmetry and space-time Killing spinors, the local brane densities involved in 1/4-BPS intersecting brane systems. In particular, we classify the possible local brane structures that have maximal (16) supersymmetries in 1/4-BPS intersecting brane backgrounds. 
Applied to BPS black holes, this classification reveals the allowed local microstructure for pure microstates.
We further use these structures with local 16 supersymmetries as building blocks to generalise to 1/8-BPS systems. Moreover, we give examples of 1/8-BPS black holes for which the local supersymmetries are compatible with the combination of different entropy-generating effects from brane interaction.
Finally, applying our classification to BPS domain walls, we illustrate how our formalism may possibly describe the local picture of the Hanany-Witten effect.

\end{abstract}
\end{adjustwidth}

\thispagestyle{empty}
\clearpage



\baselineskip=14.5pt
\parskip=3pt

\tableofcontents

\baselineskip=15pt
\parskip=3pt

\clearpage

\section{Introduction}
\label{sec:Intro}

In string theory, BPS black holes can be made out of branes and strings.
Extend along the compact (\textit{i.e.} internal) spatial directions, branes and strings of different species (intersecting branes or parallel branes) can admit collectively a supergravity description in the higher-dimensional theory. The usual technique is that of the `harmonic-function rule' \cite{Papadopoulos:1996uq,Tseytlin:1996bh,Gauntlett:1996pb}. One delocalises the branes along the compact dimensions they do not wrap, and as a result, the brane and strings of different species of branes are associated with harmonic functions, $H_I(r)$, of the same power dependence in the radial direction, $r$, in the non-compact spatial dimensions. Thus, the harmonic-function rule engineers the metric and NS and RR fields in terms of the harmonic functions, $H_I$.

Reducing to the lower-dimensional theory, the supergravity solution can describe a black hole, whose entropy depends on the number of branes of different species. The black-hole entropy can be written in the schematic form, $S\sim 2\pi \sqrt{f(N_I)}$, where the form of function $f$ depends on the number of supersymmetries that are broken. 
For example, the five-dimensional D1-D5-P black hole has three charges, which preserve together 4 of the 32 supercharges of the Type-IIB vacuum; the leading order of the entropy is of the form $S\sim 2\pi \sqrt{N_1N_5N_P}$. 

More generally, starting from a Type-II string-theory vacuum, with 32 supersymmetries, each brane/string that constitutes the black hole breaks half of the supersymmetries. The black hole thus preserves $32/2^n$ real supercharges, and is said to be $1/2^n$-BPS, with $n$ being the number of supersymmetric-compatible branes/string species.
Thus, if one adds a further species of brane/string, it looks like (at least parts of) the supersymmetries of the resulting solution will be inevitably further broken.

However, this statement is not entirely correct. It is actually possible to add branes/strings in the form of \textit{dipoles} \cite{Bena:2011uw,Bena:2022wpl} into a given supersymmetric configuration, without breaking further supersymmetries. 
An archetypal example is provided with the F1-D0 system, which consists of F1 strings wrapping a circle, $S^1$, along which there is a constant D0-brane charge. Consider a D2 brane wrapping the circle $S^1$ and a closed, topologically trivial curve, parameterised by $\psi$, in the remaining eight-dimensional spatial dimensions. The shape of the curve can be stabilised by momentum along $\psi$, and besides, the D2 brane can carry locally F1- and D0-brane charges. If one chooses exactly the right amount of local F1-, D0, and momentum-charge densities along $\psi$, the configuration can break exactly the same supersymmetries as the simple F1 and D0 charges we started with \cite{Mateos:2001qs}. Moreover, the D2 brane wraps a closed curve, so it does not introduce any global D2 charge: Along any Cartesian direction in $\mathbb{R}^8$, the positive D2 charge on one side of the curve along $\psi$ cancels with the negative D2 charge on the other side of the curve. The same happens with the momentum, $\mathrm{P}(\psi)$, which stabilises the D2 curve. Therefore, the D2 and momentum charges are local and dipolar.

With the addition of such D2-P dipoles to the F1-D0 brane charges, the resulting system does not break further supersymmetries; in fact, it has the same supersymmetries as the (nine-dimensional) F1-D0 black hole. In other words, a brane system is entitled to have complicated structure via the presence of dipoles. While the naïve brane system that has no structure backreacts into the black-hole solution in supergravity, it is natural to think that the brane system with supersymmetric-compatible dipolar excitations could describe black-hole microstates.

Thus, with this perspective, the black-hole solution, which represents a naïve brane system, is in fact an average out of the moduli space of its microstates -- more general brane systems that are supplemented by dipolar excitations. Due to the presence of the dipoles, the number of supersymmetries is increased locally, but as one moves in space, the locally preserved supercharges rotate, and so the \textit{global} supersymmetries are the same as that of the black hole. We will call this phenomenon ``local supersymmetry enhancement'' (LSE). 

In fact, it is expected that one needs to enhance the local supersymmetries to 16 supercharges in order to have pure black-hole microstates \cite{Bena:2022sge}. The reason behind this number is that pure states should be locally dualised into fundamental objects of string theory, which, like a piece of the fundamental string, preserve locally 16 supercharges. Besides, less than 16 supercharges is not enough to be a pure state. An example of this has been shown for the NS5-F1-P system (S- and T-dual to the D1-D5-P system) -- the five-dimensional black hole defined by the three-charge brane system preserves 4 supercharges. So, here, 4 supercharges is for the black hole, \textit{i.e.} the maximally mixed state. In \cite{Bena:2022sge}, it was shown that a configuration with supersymmetries enhanced up to 8 supercharges corresponds to a five-dimensional black hole with a horizon of vanishing area -- therefore, the configuration corresponds to a partially mixed state. 

This principle of enhancing the local supersymmetries of BPS brane systems is part and parcel of the Fuzzball hypothesis \cite{Bena:2022ldq,Bena:2022rna,Mathur:2005zp,Bena:2007kg,Skenderis:2008qn}, according to which there exists a basis of the black-hole microstates' Hilbert space with only horizonless microstates.  In particular, enhancing the local supersymmetries in order to find the brane-density composition of the microstructure of pure states is at the heart of the Microstate Geometries programme \cite{Bena:2013dka}, which endeavours to construct horizonless black-hole microstates in the regime of supergravity (\textit{e.g.} \cite{Bena:2016ypk,Bena:2017xbt}). 

Recently, there has been substantial advances in this regard. Indeed, the canonical examples in string theory where one could count and describe the black-hole microstates are done in the regime of very weak gravitational coupling (\textit{i.e.} one needs the 't~Hooft coupling to be small $g_s N\ll 1$) \cite{Sen:1995in,Strominger:1996sh,Dijkgraaf:1996cv,Maldacena:1997de}. 
But when one tunes the string coupling to larger values by keeping the number of branes fixed, the branes start interacting. 
And the very interest of the local supersymmetry enhancement (LSE) technology is that it takes into account of the effect of the interaction between the branes.
If one considers a microstate as a brane system at zero coupling -- \textit{e.g.} a collection of momentum-carrying, fractionated M2 strips between M5 branes \cite{Dijkgraaf:1996cv} -- the LSE of the M2-M5-P supercharges reveals what the brane system becomes once the M2 strips backreact and pull the worldvolume of the M5 branes \cite{Bena:2022wpl}, see also \cite{Bena:2022fzf,Li:2023jxb,Bena:2023rzm,Bena:2023fjx}. Thus, it instructs us of what the black-hole microstates at non-zero coupling look like.

Should this principle of enhancing the local supersymmetries to describe the microstructure of pure states be correct, it should apply to all supersymmetric black holes made of branes and strings. For each such black hole and its corresponding naïve brane system, one should find all the entire moduli space of compatible dipolar excitations with 16 local supersymmetries. One should then investigate whether this moduli space is large enough to account for a finite fraction of the black-hole entropy. 

In this work, we perform the first step of this endeavour. In Section \ref{sec:Sec2}, we first consider all the $1/4$-BPS systems, and enumerate \textit{all} the possible local supersymmetry enhancements. That is to say, given a $1/4$-BPS brane background, we list and classify all the dipoles that can exist on top of it. 
In Sections \ref{sec:Sec3} and \ref{sec:Sec3_prime}, we further use our classification of these $1/4$-BPS building blocks of brane interactions to generate local supersymmetry enhancements (LSEs) of $1/8$-BPS systems (corresponding to three-charge black holes), up to the maximal 16 local supersymmetries. In particular, we work out in Section \ref{sec:Sec3} some generalities for three-charge LSE's. Afterwards, in Section \ref{sec:Sec3_prime}, we describe in four concrete new examples of LSE for $1/8$-BPS systems and discuss their physics. We then conclude in Section \ref{sec:Concl}.

\section{Quarter-BPS building blocks: classification}
\label{sec:Sec2}

\subsection{Duality map for 1/4-BPS systems}
\label{subsec:DualityMapFor1/4-BPSSystems}

Each brane species\footnote{In this article we use the terminology brane to denote all string theory objects, i.e. not only Dp- and NS5-branes but also Kaluza-Klein monopoles (KKM), fundamental strings (F1), momenta (P) or exotic branes like the $5_2^2$-brane.} has an associated projector $\Pi$ and, equivalently, an involution $P$ describing the constraint on the Killing spinor $\epsilon$
\begin{equation}
    \Pi \epsilon \equiv \frac{1}{2} \, (1+P) \epsilon = 0.
\end{equation}
The involution is a product of gamma matrices, indicating the directions the brane extends along, and a Pauli matrix, characteristic for the type of brane. For example, $P = \Gamma^{01} \sigma_3$ is the involution of a fundamental string extended along the direction $1$, or $x^1$. A list of all involutions associated to branes in Type-II string theory can be found in the appendix \ref{sec:projectors_and_involutions_for_branes}. (For a more detailed review on these projectors/involutions, see \textit{e.g.} \cite{Smith:2002wn}.)

\begin{sidewaysfigure}[tb]
    \centering
    \includegraphics[width = \textwidth]{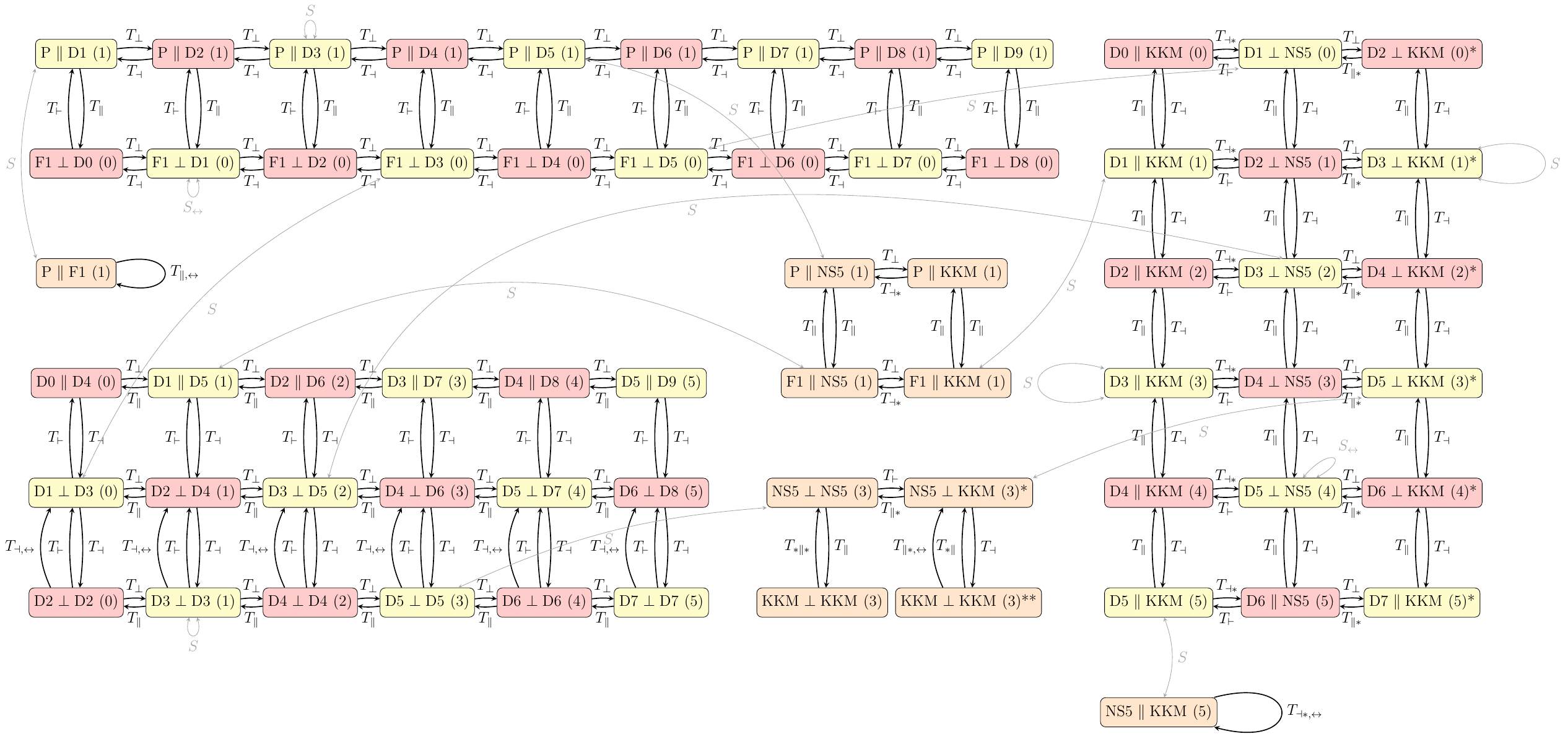}
    \caption{Duality map for standard 1/4-BPS configurations. For details or notation we refer to the text \ref{text:notation}.}
    \label{fig:DualityMap}
\end{sidewaysfigure}

\begin{figure}[tb]
    \centering
    \includegraphics[width = \textwidth]{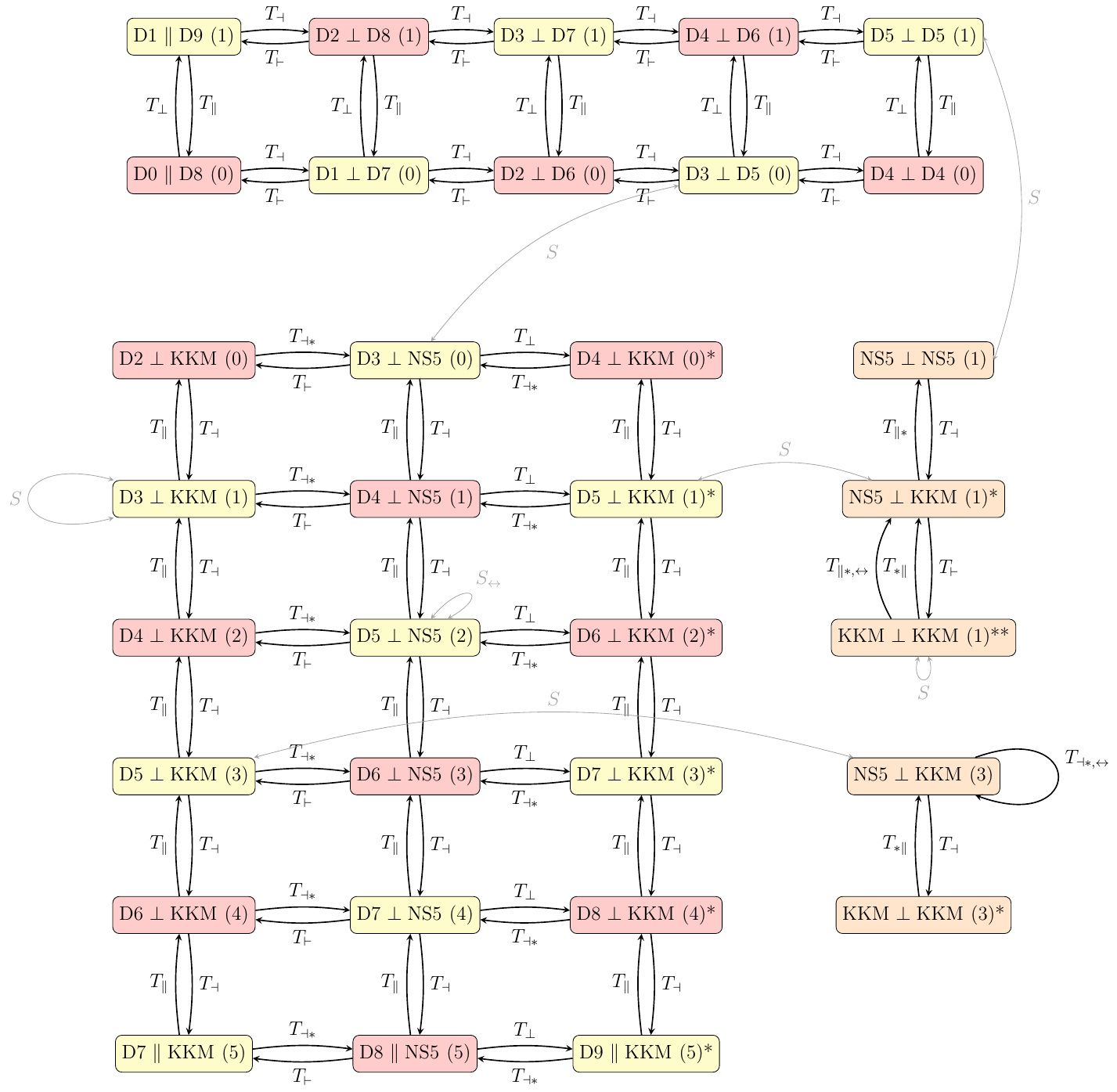}
    \caption{Duality net for non-standard 1/4-BPS configurations. For details or notation we refer to the text \ref{text:notation}.}
    \label{fig:DualityMapNonStandard}
\end{figure}

Introducing another orthogonal (or parallel) brane to the geometry, their projectors either commute or anticommute. If they commute, the projectors can be simultaneously diagonalized and we have a 1/4-BPS configuration; otherwise all supersymmetry is broken. We can find all possible 1/4-BPS configurations for flat backgrounds in this way and connect them via various dualities. The result is a duality map with all 1/4-BPS configurations. 
It has two completely disconnected components, the \textit{standard} configurations, depicted in Figure \ref{fig:DualityMap}, and the \textit{non-standard} configurations, depicted in Figure \ref{fig:DualityMapNonStandard}.\footnote{We use the same terminology as in \cite{Boonstra:1998yu}, see also \cite{Gauntlett:1997cv}.}
Both connected components have a similar substructure. The 1/4-BPS configurations are grouped in blocks that are closed under (allowed) T-dualities, while the S-dualities connect these blocks. It is not possible to go from a standard configuration to a non-standard one via T- and S-dualities, even when introducing exotic branes. This will be further discussed in subsection \ref{subsec:RemarkOnExoticBranes}.

\subsubsection*{Notation}
\label{text:notation}
This part clarifies the notation used in the diagrams in this section. It can be skipped and returned to, if the notation is not intuitively clear.

We will denote a 1/4-BPS configuration consisting of the branes $A$ and $B$ by $A_{123} \perp B_{14567} (1)$, or by $A_{123} \parallel B_{12345} (3)$, if one brane is completely inside the other. The bracket at the end denotes the number of common directions. Further stars * at the end denote that the special direction of one or both KKMs are inside the other brane. 
The indices refer to the directions along which the brane is extended, but may be omitted to shorten the notation.%
\footnote{Usually the special direction of the KKM is separated from the indices by a semicolon, as in $\KKM[12345;6]$. We decided to omit the special direction everywhere, as different special directions are possible.}

A T-duality along a direction that is parallel / perpendicular to both branes is denoted by $T_{\parallel}$ / $T_{\perp}$. $T_{\vdash}$ denotes that the direction is parallel to the first brane and perpendicular to the second, and the other way around for $T_{\dashv}$. 
A star $*$ before (or after) the symbol $\parallel$ or $\dashv$ denotes a transformation along the special direction of the first (or second) KKM. If a duality interchanges the type of the first and second brane, we will denote this by an arrow $\leftrightarrow$. 

A yellow background colour stands for a configuration in Type IIB, a red one for Type IIA. Orange configurations exist in both. A self-duality of these configurations, i.e. to go from a Type IIA to a Type IIB theory or vice versa, is usually not explicitly denoted in the duality map.

For the diagrams that include the possible glues for a local supersymmetry enhancement (see section \ref{subsec:ApplicationToTheDualityMap}), we split the nodes in two parts. The first part stays the same, while the second part is a list of all possible glues that fit the scheme presented below. Orange nodes, i.e. nodes with main branes that exist in both Type II theories, have three parts where the second part lists all glues in Type IIA theory, and the third in Type IIB theory.

\clearpage
\subsection{A general ansatz for local supersymmetry enhancement}

As indicated in the Introduction \ref{sec:Intro}, in order to formulate generalities for local supersymmetry enhancement (LSE) between multiple species of branes, we first focus on local supersymmetry enhancements between two species of branes. 
By this, we mean that globally the configuration is 1/4-BPS, i.e. preserves 8 of the 32 supersymmetries, as it would be expected from two branes with commuting projectors. However, if we zoom in on a point on the worldvolume of the `enhanced' configuration, (\textit{i.e.} with the addition of the dipole charges that we call \textit{glues}), the amount of supersymmetries is enhanced to 16 supercharges. 

We start with a 1/4-BPS configuration composed of two brane species -- referred to as \textit{main-brane species} or \textit{main branes} -- characterised by the involutions $P_1$ and $P_2$, \textit{i.e.} $(P_1)^2=(P_2)^2=1$. Furthermore, we will add two other brane species, referred to as \textit{glues}, with the involutions $Q_1$ and $Q_2$. It will be apparent, later, that those can be interpreted as dipoles or oscillating charge densities with vanishing net charge.

Let us remark that in our terminology, a \textit{main-brane species} denotes a type of brane carrying charges along some predetermined directions. For example one can consider a D2-brane charge along the directions $012$ and D2-brane charge along $034$ as \textit{main branes}, and D2-brane charges along $013$ and along $024$ as \textit{glues}.

We use the method described in \cite{Bena:2022wpl,Li:2023jxb}. We make an ansatz for a projector $\Pi$ for the whole system
\begin{equation}
    \Pi \equiv \frac{1}{2} \left( 1 + \alpha_1 (x) \, P_{1} + \alpha_2 (x) \, P_{2} + \beta_1 (x) \, Q_{1} + \beta_2 (x) \, Q_{2} \right),
\end{equation}
with the coefficients $\alpha_i$ and $\beta_i$ describing the ratio of the charge density of the associated object to the overall mass density, see \cite{Bena:2022wpl,Li:2023jxb}.

Then, we demand that it splits into a sum of the two projectors $\Pi_i = \frac{1}{2} (1+P_i)$ of the main branes to ensure that eight global supersymmetries are preserved.\footnote{We could also demand a split into $\Pi_i = \frac{1}{2} (1-P_i)$. The solution remains the same apart from a relative sign in front of the main branes' coefficients. This solution can also be generated by the transformation $P_i \rightarrow - P_i$.} Together with the projector condition, we get a set of two (matrix-valued) equations,
\begin{align}
    \Pi^2 &= \Pi \, \label{eq:projector_condition} ,\\
    \Pi &= f_1 (x) \, \Pi_{1} + f_2 (x) \, \Pi_{2} \, , \label{eq:ansatzforglobalsupersymmetry}
\end{align}
where $f_1$ and $f_2$ are matrix-valued functions on the worldvolume of the whole system (bound state).

We take the ansatz
\begin{subequations}
\label{eq:ansatz_P1withQ1}
\begin{align}
    f_1 &= \alpha_1 + \beta_1 \, Q_1 P_1 \, , \\
    f_2 &= \alpha_2 + \beta_2 \, Q_2 P_2 \, .
\end{align}
\end{subequations}
This associates the first glue with the first main brane and the second glue with the second main brane. This association can also be switched, see \cite{Li:2023jxb}. Often there is an exterior reason to associate one main object with one specific glue, \textit{e.g.} being the same type of brane when dualised to M-theory, but from the LSE formalism point of view, this association is arbitrary.

With the ansatz \eqref{eq:ansatz_P1withQ1}, we find the following conditions on the coefficients,
\begin{subequations}
\begin{align}
    1 &= \alpha_1 + \alpha_2 \, , \label{eq:BPScondition}\\
    \alpha_1 \alpha_2 &= - \eta \, \beta_1 \beta_2 \, , \label{eq:anticommutatorEquation}\\
   \beta_1 &= - \beta_2 \, \eta \, , \label{eq:PQQP} \\
    0 &= \alpha_1 \alpha_2 \{ P_1 , P_2 \} + \beta_1 \beta_2 \{ Q_1 , Q_2 \} \nonumber\\
    &\quad + \sum_{i,j} \alpha_i \beta_j \{ P_i , Q_j \} \, , \label{eq:2chargeAnticommutatorConstraint}
\end{align}
\end{subequations}
where 
\be \label{def_eta}
\eta \equiv P_1 Q_1 Q_2 P_2 \, . 
\ee

The first line is the usual BPS condition. As the coefficients are charge-mass ratios, we can rewrite it into 
\begin{equation}
    M = Q_1 + Q_2 \, .
\end{equation}
The second equation, \eqref{eq:anticommutatorEquation}, is a quadratic equation relating the charges of the glues with the charges of the main branes: The brane densities of the main branes are related to the brane densities of the glues. The third equation, \eqref{eq:PQQP}, tells that the absolute value of charges of the glues are constrained to be equal everywhere in the bound state. 

In equation \eqref{eq:PQQP}, the key condition on a 2-charge local supersymmetry enhancement emerges. As the coefficient are numbers, the product of involutions $\eta$ has to be proportional to the identity. 
By looking at the determinant of $\eta$, we can further prove that $\eta$ is a sign, $\eta = \pm 1$. 
It controls whether the glues are two branes, two anti-branes, or a brane and an anti-brane, given the orientation of space one has previously fixed.
Furthermore, by inserting $1 = P_1^2$ (and then $1 = Q_1^2$, etc.) we can show that $\eta$ is invariant under circular shifts of its constituent involutions,
\begin{equation}
    \eta = \eta P_1^2 = P_1 \eta P_1 = P_1 P_1 Q_1 Q_2 P_2 P_1 = Q_1 Q_2 P_2 P_1 \, .
\end{equation}

The (anti-)commutation relations between the involutions of the branes are heavily constrained by $\eta$. The main branes commute to preserve some global supersymmetry. All other relations follow from the relation between $Q_1$ and $Q_2$. This can be proven by cycling the involutions through $\eta$, and computing $\eta^2 = 1$. This gives the following two possibilities:
\begin{itemize}
    \item The two glues commute, but main branes and glues anticommute with each other, or 
    \item all involutions $P_i$ and $Q_j$ commute.
\end{itemize}

Both options have different consequences on how we can satisfy equation \eqref{eq:2chargeAnticommutatorConstraint}.
The first and more interesting option gives the following relations:
\begin{subequations}
\begin{align}
    [ P_1 , P_2 ] &= [ Q_1 , Q_2 ] = 0 \label{eq:commutatorCondition} \\
    \{ P_i , Q_j \} &= 0 \label{eq:anticommutatorCondition}
\end{align}
\end{subequations}
In order to satisfy equation \eqref{eq:2chargeAnticommutatorConstraint}, we have to cancel the two non-vanishing anticommutators and their coefficients with each other,
\begin{equation} 
    0 = \alpha_1 \alpha_2 \{ P_1 , P_2 \} + \beta_1 \beta_2 \{ Q_1 , Q_2 \} \, . 
\end{equation}
We can do so by using the quadratic equation \eqref{eq:anticommutatorEquation}, the (anti-)commutation rules of the involutions and $\eta = Q_2 P_2 P_1 Q_1 = \pm 1$.
\begin{align}
    \alpha_1 \alpha_2 \, \{ P_1 , P_2 \} &=- \beta_1 \beta_2 \, \eta \, (P_1 P_2 + P_2 P_1) \nonumber\\
    &=- \beta_1 \beta_2 \, (P_1 \eta P_2 + P_2 \eta P_1) \nonumber\\
    &=- \beta_1 \beta_2 \, (Q_1 Q_2 + P_2 Q_2 P_2 P_1 Q_1 P_1) \nonumber\\
    &=- \beta_1 \beta_2 \, (Q_1 Q_2 + Q_2 Q_1) = - \beta_1 \beta_2 \, \{ Q_1 , Q_2 \} \, . \label{eq:cancellationanticommutators}
\end{align}

Let us consider the second option, where all involutions commute. We can still use the same line of reasoning to cancel the two anticommutators above, but further anticommutators appear in equation \eqref{eq:2chargeAnticommutatorConstraint}.
\begin{equation}
\begin{split}
    0 =& \alpha_1 \beta_1 \{ P_1 , Q_1 \} + \alpha_1 \beta_2 \{ P_1 , Q_2 \} \\
    &+ \alpha_2 \beta_1 \{ P_2 , Q_1 \} + \alpha_2 \beta_2 \{ P_2 , Q_2 \} 
\end{split}
\end{equation}
To cancel those anticommutators with each other in the same way as above, we have to introduce two new equations,
\begin{subequations}
\begin{align}
    \alpha_1 \beta_1 &= - \eta \, \alpha_2 \beta_2 \\
    \alpha_1 \beta_2 &= - \eta \, \alpha_2 \beta_1 \, .
\end{align}
\end{subequations}
While there is a solution to these equations ($\alpha_i = \beta_j = \pm 1/2$), it does not have any degrees of freedom. As we cannot vary the coefficients in space, this is a 1/2-BPS configuration with no notion of local supersymmetry. 

Returning to the first option, it does not introduce new equations and only has to satisfy the equations \eqref{eq:BPScondition} to \eqref{eq:PQQP}. They are solved by
\begin{align}
    \alpha_1 &= \cos^2(\theta) \nonumber\\
    \alpha_2 &= \sin^2(\theta) \nonumber\\
    \beta_1 &= \kappa \, \sin(\theta) \cos(\theta) \nonumber\\
    \beta_2 &= - \eta \, \kappa \, \sin(\theta) \cos(\theta) \,, 
\end{align}
where we have introduced the sign $\kappa = \pm 1$, which controls whether the first glue is a brane or an antibrane.
In contrast to the constant solution, we find a parameter $\theta = \theta(x)$ that can vary in space. 

To summarize, we have found two conditions on the involutions $P_i$ and $Q_j$,
\begin{align}
    \eta \equiv P_1 Q_1 Q_2 P_2 & = \pm 1 \, , \label{eq:etaCondition}\\
    [ P_1 , P_2 ] & = 0 \, , \tag{\ref{eq:commutatorCondition}} \\
    \{ P_1 , Q_1 \} & = 0 \, . \tag{\ref{eq:anticommutatorCondition}} 
\end{align}
If these conditions are fulfilled, there exists a solution with one degree of freedom for the charges of the involved branes where the amount of supersymmetries is doubled when zooming in on a specific point of the worldvolume of the branes.


\subsection{Application to the duality map}\label{subsec:ApplicationToTheDualityMap}

With the conditions on local supersymmetry enhancement (LSE), we can apply them to the duality maps in Figure \ref{fig:DualityMap} and \ref{fig:DualityMapNonStandard} in order to find 2-charge LSEs for Type-II string theories on a flat ten-dimensional background.

Take a given 1/4-BPS configuration. This fixes the involutions $P_1$ and $P_2$ of the main brane species and automatically satisfies the condition \eqref{eq:commutatorCondition}. As listed in the Appendix \ref{sec:projectors_and_involutions_for_branes}, the involutions $P_i$ and $Q_j$ consist of a product of Gamma matrices tensored with a Pauli matrix. The Gamma matrices indicate the directions along which the branes are extended, and the Pauli matrix determines the type of object/brane. The condition $\eta = \pm 1$ can be split into two conditions on the corresponding vector spaces of the Gamma/Pauli matrices.
This will narrow down the possible configurations to a handful. Afterwards, we check whether the anti-commutator $\{P_1,Q_1\}$ vanishes to confirm a local supersymmetry enhancement.

We illustrate this procedure for the \W-\D{1} configuration extended along the direction $x^1$. In the Table \ref{tab:paulimatricesofbranes} below, we can read off the Pauli matrices inside the projectors of the branes, $\sigma_{\W} = 1$ and $\sigma_{\D{1}} = \sigma_1$. Since the product of all involved Pauli matrices should be proportional to 1, we can choose for the glues either one from the first two columns each, or from the last two columns, \textit{e.g.} \KKM-\D{5} or \F-\D{3}.

\begin{table}[htb]
    \centering
    \begin{tabular}{c|c c c c}
        $\sigma$    & $1$   & $\sigma_1$    & $i \sigma_2$ & $\sigma_3$ \\\hline
                    & \W    & \D{2}         & \D{0}         & \F        \\
        Branes (IIA)& \NS   & \D{6}         & \D{4}         & \KKM      \\
                    &       &               & \D{8}         &           \\\hline
                    & \W    & \D{1}         & \D{3}         & \F        \\
        Branes (IIB)& \KKM  & \D{5}         & \D{7}         & \NS       \\
                    &       & \D{9}         &               &           
    \end{tabular}
    \caption{Pauli matrices in the involutions of branes in string theory.}
    \label{tab:paulimatricesofbranes}
\end{table}

The condition on the Gamma matrices is analogous; their product should be proportional to 1. Since the $\Gamma$-matrices of the main branes already cancel each other (up to a sign), we get the condition $\Gamma_{Q_1} \propto \Gamma_{Q_2}$. This is only the case if the glues have the same dimensions, and are extended along the same dimensions. There is no such combination in the last two columns of Table \ref{tab:paulimatricesofbranes}, so we can disregard this case. In the first two columns, we can find $\W-\D{1}$ and $\KKM-\D{5}$. For both configurations, we can extend them along the direction $x^1$ of the main branes, or a transverse direction like $x^9$. 
The last condition for a local supersymmetry enhancement is $\{P_1,Q_1\} = 0$ and $[Q_1 , Q_2] = 0$. This excludes the configurations parallel to the string, as seen below.
\begin{align*}
    \{ P_{\W[1]} , P_{\W[1]} \} &= 2 \neq 0 \\
    \{ P_{\W[1]} , P_{\W[9]} \} &= 0 \\
    \{ P_{\W[1]} , P_{\KKM[12345]} \} &= 2 \, \Gamma^{2345} \neq 0 \\
    \{ P_{\W[1]} , P_{\KKM[56789]} \} &= 0 \\
\end{align*}

\begin{figure}[htb]
    \centering
    \includegraphics[width = 0.8 \textwidth]{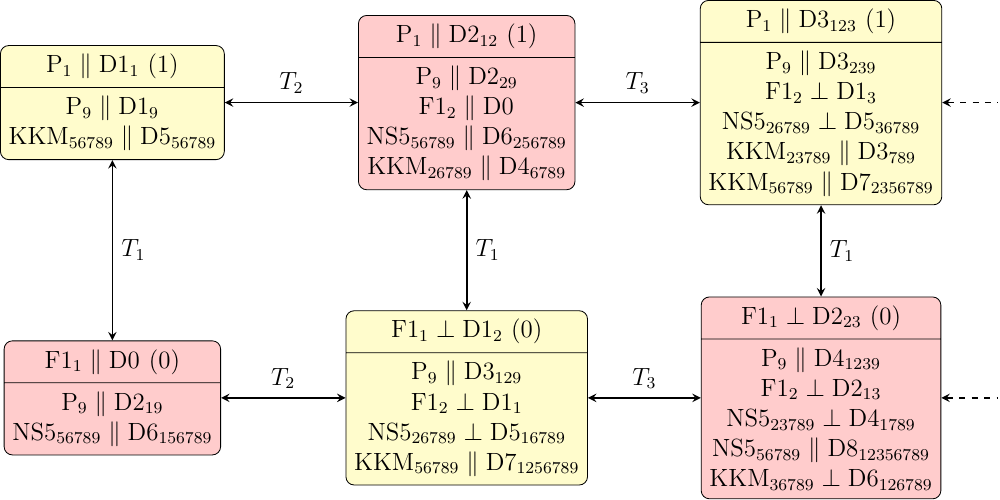}
    \caption{A part of the duality map with possible local supersymmetry enhancements. For details, we refer to the text.}
    \label{fig:P-Dp-Configurations-Shortened}
\end{figure}

To summarize, we find two possible local supersymmetry enhancements for the 1/4-BPS configuration $\W[1]-\D[1]{1}$. We can use $\W[9]-\D[9]{1}$ or $\KKM[56789]-\D[56789]{5}$ as glues. This is represented by the first node in Figure \ref{fig:P-Dp-Configurations-Shortened}. The top part of the node denotes the main branes, while the second lists all possible glues for a local supersymmetry enhancement. We applied this procedure to every 1/4-BPS configuration in the duality map from Section \ref{subsec:DualityMapFor1/4-BPSSystems} to find all LSEs that fit the scheme presented above. The resulting diagram is very large, so we split it into the different blocks connected by S-dualities and included them in the Appendix (Figures \ref{fig:P-Dp-Configurations} to \ref{fig:Dp-NS5-NonStandard-Configurations-2}). 

\subsubsection*{The inverse supertube transition}\label{sssec:TheInverseSupertubeTransition}

Upon further inspection of all the local supersymmetry enhancements, we can find an inversion symmetry between the main branes and glues: If $A \perp B$ admit an LSE by adding the glues $C \perp D$, then we can also construct a different configuration with $C \perp D$ as main branes and enhance the system to 16 local supersymmetries by adding $A \perp B$ as glues.
These are two distinct configurations, as the solutions for the coefficients $\alpha_i$ and $\beta_j$ are also different. The main point is that the existence of local supersymmetry enhancement for one configuration implies the same for the other. This is easily proven by considering the conditions on an LSE in equations \eqref{eq:etaCondition} and \eqref{eq:anticommutatorCondition} again. The (anti-) commutator relations are symmetric under exchanging the main branes with the glues, and for the condition on $\eta$ we can use the invariance under circular shifts and the commutators between the main branes and glues to prove the symmetry,
\begin{equation}
    \pm 1 = \eta \equiv P_1 Q_1 Q_2 P_2 = Q_1 Q_2 P_2 P_1 = Q_2 Q_1 P_1 P_2 = Q_1 P_1 P_2 Q_2 \, .
\end{equation}

Let us illustrate this symmetry for the supertube transition, where an $\F$-string with constant $\D{0}$ charge is puffed out into a tubular $\D{2}$ configuration carrying angular momentum. 
For the 'inverse' transition, we can start with a $\D[y1]{2}$ brane along some direction $y$ and $1$ with momentum along $1$, and enhance the configuration by adding internal $\F[y]$ and $\D{0}$ dipole charges along $y$. From an M-theory point of view, the two configurations are almost identical. They correspond to an $\M{2}$ brane with momentum performing wiggles in either the $y$, or the eleventh dimension.

\subsection{Discussion of the local supersymmetry enhancements} \label{subsec:DiscussionOfTheLocalSupersymmetryEnhancements}

The D2-P supertube with F1-D0 charges is one example of a well-known configuration with an enhanced local supersymmetry. Naturally, this local supersymmetry enhancement can be found in the extensive list of glues for 1/4-BPS system in the appendix, explicitly in Figure \ref{fig:P-Dp-Configurations}.  We want to specifically highlight this and some other local supersymmetry enhancements we found in the previous subsection.

In this part and in the following, we denote a local supersymmetry enhancement (LSE) of the main branes $A\perp B$ involving the glues $C$ and $D$ by:
\be
\[ \, A\perp B \, \] \;
\rightarrow \;
\( \, C \,,\, D \, \) \, .
\ee
Then, the original F1-D0 supertube transition \cite{Mateos:2001qs,Emparan:2001ux} involves the LSE
\be \label{eq:supertube_transition_F1D0}
\[ \, \F[1] \parallel \D{0} \; (0) \, \] \;
\rightarrow \;
\( \, \D[1\psi]{2} \,,\, \W[\psi] \, \) \, .
\ee
$\F$ strings with constant $\D{0}$ charge are puffed out into a $\D{2}$ tubular configuration carrying angular momentum. Along $\psi$, the $\D{2}$-brane follows an arbitrary closed profile, which is stabilised due to the presence of the angular momentum, $\W[\psi]$. 
A higher-dimensional analogue of \eqref{eq:supertube_transition_F1D0} is:
\be
\[ \, \F[1] \parallel \D[2\cdots]{p} \; (0) \, \] \;
\rightarrow \;
\( \, \D[12\cdots\psi]{(p+1)} \,,\, \W[\psi] \, \) \,
\ee
These transitions can also be found in Figure \ref{fig:P-Dp-Configurations-Shortened}, a part of the whole block with $\W$-$\D{p}$ and $\F$-$\D{p}$ configurations displayed in Figure \ref{fig:P-Dp-Configurations} that we included here for illustration purposes.

This kind of supertube transition, where the 1/4-BPS brane system puffs out into a bound state of one higher dimension, also has T- and S-dual avatars in other blocks. For instance, the transition
\be \label{D0D4_supertube}
\[ \, \D{0} \parallel \D[1234]{4} \; (1) \, \] \;
\rightarrow \;
\( \, \NS[1234\psi] \,,\, \W[\psi] \, \) \,
\ee
or its T-dual along $y$
\be \label{D1D5_supertube}
\[ \, \D[y]{1} \parallel \D[y1234]{5} \; (1) \, \] \;
\rightarrow \;
\( \, \KKM[1234\psi;y] \,,\, \W[\psi] \, \) \,
\ee
The transitions \eqref{D0D4_supertube} and \eqref{D1D5_supertube} can be understood as a point-like brane system in the non compact directions (respectively $\mathbb{R}^5$ and $\mathbb{R}^4$) that is puffed out into a closed curve along $\psi$ in the non-compact directions. The transition \eqref{D1D5_supertube} is at the core of the construction of the Lunin-Mathur and LMM geometries \cite{Lunin:2001fv,Lunin:2002iz}.

There is a way to add to a brane a parallel momentum charge consistently with 16 local supersymmetries. Take, for example, a $\D{p}$ brane with an extension along $y$; adding a parallel momentum along $y$ admits the following LSE:
\be
\[ \, \W[y] \parallel \D[y1\cdots]{p} \; (1) \, \] \;
\rightarrow \;
\( \, \W[\psi] \,,\, \D[\psi1\cdots]{p} \, \) \,,
\ee
and similarly, with a fundamental string:
\be \label{F1-P_transition}
\[ \, \W[y] \parallel \F[y] \; (1) \, \] \;
\rightarrow \;
\( \, \W[\psi] \,,\, \F[\psi] \, \) \,.
\ee
The above two transitions tell that one can realise microscopically a parallel momentum along a brane/string, by having the brane or the string taking a profile in the transverse directions (along $\psi$), in the $(y,\psi)$ plane. Because of supersymmetry, the profile is independent of one of the null coordinates, $u$, of $S^1_y$ and thus travels at the speed of light along its orthogonal null-like coordinate, $v$: Momentum is hence produced by the profile. 
Such configurations admit a supergravity description which depends on the shape of the profile, see \cite{Dabholkar:1995nc}.

Another well-understood geometry with 16 local supersymmetries is the Callan-Maldacena spike, or BIon spike \cite{Callan:1997kz}:
\be
\[ \, \D[y]{1} \perp \D[123]{3} \; (0) \, \] \;
\rightarrow \;
\( \, \D[1]{1} \,,\, \D[y23]{3} \, \) \, .
\ee
Instead of a string ending on a $\D{3}$ brane perpendicularly, the string pulls on the worldvolume of the $\D{3}$ brane forming a smooth spike. 
To retain the rotational symmetry of the $\D{3}$ branes, we interpret the distinct direction $x^1$, along which the glue $\D[1]{1}$ extends, as the radial direction of the $\D{3}$ brane.
Close to the would-be endpoint of the string, the effect of the glues becomes apparent. The $\D{3}$ brane tilts into the direction $y$ of the string, and vice versa, the string tilts towards the $\D{3}$ brane. This creates a spike in the $\D{3}$ brane, where the string can be recovered at infinite distance \cite{Callan:1997kz}. Placing another $\D{3}$ brane at a finite distance, it is shown in \cite{Hashimoto:1997px} that the radius of the tubes coming from the branes vanishes linearly in the middle of the two $\D{3}$ branes, forming an X-like cross-section instead of an hourglass shape.

A similar configuration can be achieved by using a fundamental string instead of a D-string, or by using higher dimensional branes, 
\be \label{Dp-F1_with_Dp-F1}
\[ \, \D[x2\dots p]{p}\perp \F[y] \; (0) \, \] \;
\rightarrow \;
\( \, \D[y2\dots p]{p} \perp \F[x] \, \) \, .
\ee
Therefore, the transition \eqref{Dp-F1_with_Dp-F1} can be interpreted as a Callan-Maldacena transition, in general dimensions, between a D$p$ brane and a fundamental string ending orthogonally on it.

A special case of \eqref{Dp-F1_with_Dp-F1} is the intersection of a D-string with a fundamental string, 
\be \label{D1-F1_transition-recombination}
\[ \, \D[x]{1}\perp \F[y] \; (0) \, \] \;
\rightarrow \;
\( \, \D[y]{1} \perp \F[x] \, \) \, ,
\ee
for which we propose also an other interpretation, in terms of a pair of intersecting D1 branes with electric flux corresponding to F1-string charge \cite{Hashimoto:2003pu}.
The authors of \cite{Hashimoto:2003pu} consider two intersecting $(\pm p,1)$-strings, with the intersecting angle determined by $p$. There is an orthogonal basis that we parameterise by the coordinates $(x,y)$, such that along $x$, one measures a net charge of two D1 branes and that along $y$, one measures a net flux charge of F1 strings. Then, there is a deformation preserving supersymmetry that disconnects the D1 branes into two D1 branches with local F1-flux charges with a hyperbolic shape. In our language, this is the transition \eqref{D1-F1_transition-recombination}: the local brane densities on the two hyperbolic branches involve not only that of the main-brane charges, but also that of the glues. However we do not know yet whether the transition \eqref{Dp-F1_with_Dp-F1} could also be interpreted as the recombination of D$p$ branes.


There are many more configurations. Most of them involve higher-dimensional branes, $\NS$-branes, $\KKM$s or $\D{p}$ branes, to puff out in one or more directions.  But we do want to highlight two specific LSEs with only internal excitations, that to our knowledge did not appear in the literature before. The first one is
\be
\[ \, \F[y] \perp \D[12345]{5} \; (0) \, \] \;
\rightarrow \;
\( \, \D[y]{1} \,,\, \NS[12345] \, \) \, .
\ee
Apart from the (inverse) supertube transition and the higher-dimensional Callan-Maldacena spikes, there are only very few configurations in the $\F$-$\D{p}$ block of the duality map whose glues do not use extra dimensions. This is one of them, whose ingredients remind of $(p,q)$-strings and $(p,q)$-fivebranes. Further studies will hopefully shed more light on this.

The second configuration is an enhancement of a $\D{4}$ brane with constant $\D{0}$ charge by using two orthogonal $\D{2}$ branes inside the $\D{4}$.
\be \label{D0-D4_with_D2-D2}
\[ \, \D{0} \parallel \D[1234]{4} \; (0) \, \] \;
\rightarrow \;
\( \, \D[12]{2} \,,\, \D[34]{2} \, \) \, .
\ee
This looks like a D0-D4 instanton which admits non-vanishing local D2-D2 charges, with no net D2-D2 charges. However, revealing the dipolar/multipolar D2-D2 charges of such an instanton is not trivial, since the instanton field strength $F$ is traceless. This will hopefully be reported soon in \cite{Dulac:2024}.
We will later use the T-dual of the transition \eqref{D0-D4_with_D2-D2} along another direction $y$ to enhance the $\D{1}$-$\D{5}$-$\W$ system with internal dipoles in Section \ref{subsec:D1-D5-P-InternalDipoles}.

\vspace{1em}
At this stage, it is unclear to us whether the microscopic degeneracy of each 1/4-BPS intersecting brane system is necessarily explained by bound states involving the presence of the allowed glues. 
What the local supersymmetry enhancement reveals is that there exists, potentially, some transitions that are kinematically allowed from a supersymmetry point of view. An example we can give is the transition 
\be \label{D2-D2_with_D2-D2}
\[ \, \D[12]{2} \perp \D[34]{2} \; (0) \, \] \;
\rightarrow \;
\( \, \D[13]{2} \,,\, \D[24]{2} \, \) \, .
\ee
We think that the basic-level interpretation of this transition is the statement that: there is a way to connect a piece of D2 brane living in the directions 12 with another piece of D2 brane in directions 34, compatible with supersymmetry. In fact, the D2 brane that makes the connection can take an arbitrary holomorphic shape in terms of the complex coordinates $(x_1+ix_2,x_3+ix_4)$ \cite{Lunin:2008tf}.
Now, there is a second level of interpretation is related to the microscopic degeneracy of the 1/4-BPS brane intersection: That the degeneracy of the black hole corresponding to the D2-D2 system on $T^4$ or $K3$ is explained by D2 branes wrapping holomorphic curves in $T^4$ or $K3$. What we are unsure of is that this second-level interpretation should apply for all 1/4-BPS intersections. 

Similarly, there may be a second-level interpretation for the non-standard configurations (Fig. \ref{fig:DualityMapNonStandard}), which admit the glues in Figures \ref{fig:Dp-Dq-NonStandard-Configurations} to \ref{fig:Dp-NS5-NonStandard-Configurations-2}. The main branes of the non-standard configurations mostly occupy together 8 spatial dimensions (sometimes 9), so (when they are not space-filling) they would lead to domain walls rather than black holes. When we have one spatial dimension left (the dimension may have to be found in M theory), we can apply a similar logic to what we did for black holes: By delocalising the branes in the internal dimensions, the brane system can source a domain-wall solution in the non-compact dimensions \cite{Lust:2022lfc}. Domain walls do not have an entropy in the proper sense; however one can associate a central charge to it, if the near-brane limit admits a AdS geometry \cite{Boonstra:1998yu,deBoer:1999gea,KKLT_ex_nihilo}. Then, the central charge measures the number of degrees of freedom of the brane system constituting the domain wall. Contrary to standard brane intersections giving rise to black-hole solutions, many of the non-standard brane intersections giving rise to domain walls admit a Hanany-Witten-type transition; therefore, degrees of freedom on the domain wall can come from this non-perturbative effect \cite{KKLT_ex_nihilo}.

We are unsure whether the degrees of freedom of all such domain walls must be captured by bound states admitting local supersymmetry enhancements.
However, at the basic level, some transitions involving the glues we list for non-standard configurations can be kinematically realised. Take for example the transition
\be \label{NS5-NS5_4cycle}
\[ \, \NS[1234 9] \perp \NS[5678 9] \; (1) \, \] \;
\rightarrow \;
\( \, \NS[1235 9] \,,\, \NS[4678 9] \, \) \, ,
\ee
in Fig. \ref{fig:NS5-KKM-NonStandard-Configurations-reordered} or its D5 cousin in Fig. \ref{fig:Dp-Dq-NonStandard-Configurations}. It is indeed known that five-branes can wrap holomorphic 4-cycles or special Lagrangian 4-cycles, inside a Calabi-Yau fourfold -- and act as domain walls in M theory \cite{Lust:2022lfc,KKLT_ex_nihilo}. At the self-intersection points, the intersection is locally a $\NS[1234 9] \perp \NS[5678 9]$. The holomorphic or special Lagrangian 4-cycle connecting the two pieces of orthogonal NS5 branes will involve the NS5-NS5 glues of the form \eqref{NS5-NS5_4cycle} -- this is really similar to the picture in \eqref{D2-D2_with_D2-D2}. 

Note that after dualising the NS5-NS5 intersection to the D0-D8 frame, the transition \eqref{NS5-NS5_4cycle} gets mapped to
\be \label{D0-D8_with_D2-D6}
\[ \, \D{0} \parallel \D[12345678]{8} \; (0) \, \] \;
\rightarrow \;
\( \, \D[12]{2} \,,\, \D[345678]{6} \, \) \, .
\ee
This looks like a D0-D8 version of \eqref{D0-D4_with_D2-D2}. This LSE may play a role in describing dynamics of the D0-D8 system, in the context of the Hanany-Witten effect, see Subsection \ref{ssec:D0-D8-F1_Hanany-Witten}.


\subsection{Main branes and glues in the context of black-hole physics}
\label{subsec:dofs_ontopof_background}

The black-hole solution in supergravity results in a brane configuration that is delocalised in the internal dimensions, and point-like in the non-compact space-like dimensions. Pure microstates should evade either the delocalisation in internal dimensions, or the fact of being point-like in the non-compact spacelike dimensions. As such, we classify the glues into two categories; internal glues, that only extend along the compact directions of the system, and external glues, that expand into at least one non-compact space-like dimension.

The duality map with all possible glues is the first step towards finding the microstructure degrees of freedom that are allowed for various duality frames. When examining one specific duality frame, we provide all glues with which the configuration can be enhanced to 16 local supersymmetries. These glues then again give a picture of the black-hole microstructure in this duality frame. 

One can think of the main branes as what determines the geometric background of the black hole and its main charges. Then, some glues can be viewed as a layer on top of this background, describing various excitations of the system. An illustration of this is the $\D[1]{1}$-$\D[234]{3}$ system compactified on a torus $T^4_{1234}$. The main branes define the black-hole charges that one can measure in supergravity; when the branes are delocalised in the compact dimensions, the main branes source the geometry of the black hole. The glues $\D[2]{1}$-$\D[134]{3}$ enable the (localised) orthogonal main branes to form Callan-Maldacena spikes. The entropy is then generated by a fractionation of these spikes, where one microstate corresponds to a specific distribution of the $\D{1}$-strips forming the spikes. This is one example of internal excitations of the $\D{1}$-$\D{3}$ system. Of course, it is also possible to use external glues, like $\NS[1234 \psi]$-$\W[\psi]$, which extend along an arbitrary curve $\psi$ in the external dimensions $\mathbb{R}^4$.

\begin{figure}[h]
    \centering
    \includegraphics[width = \textwidth]{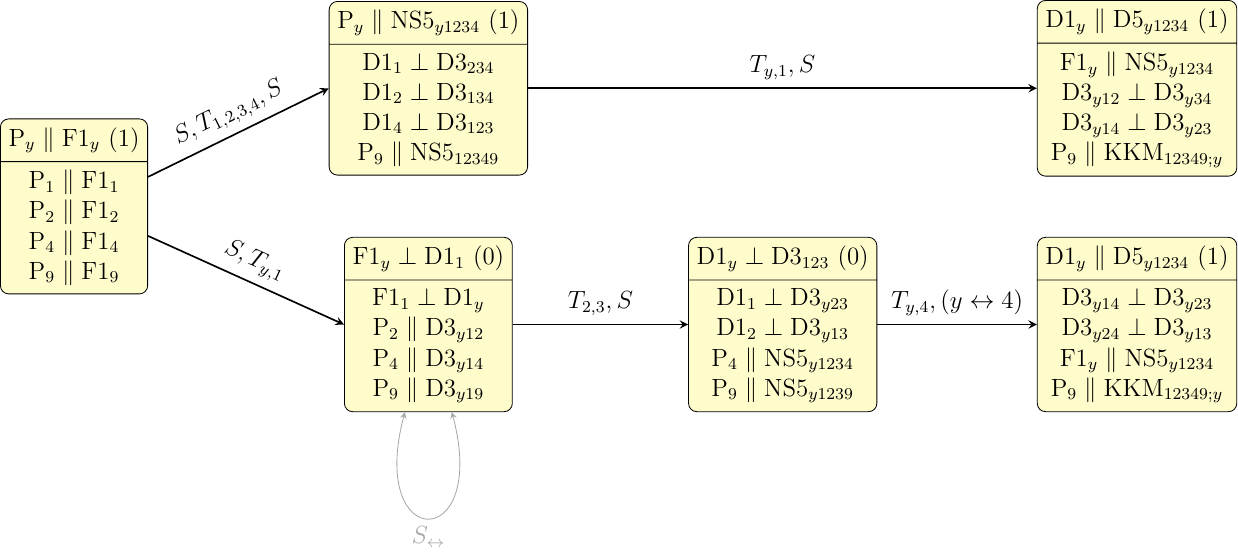}
    \caption{Illustration of two paths from a fundamental string with momentum to the $\D{1}$-$\D{5}$ intersection. The top path is used in \cite{Kanitscheider:2007wq} and \cite{Mathur:2005zp}, and compared in this chapter with the bottom path. The roles of the two main branes can be switched by inserting an S-duality when arriving at the $\F$-$\D{1}$ intersection.}
    \label{fig:PathsFromF1PToD1D5}
\end{figure}

The microstructure of a black hole with classically vanishing horizon, consisting of a fundamental string with momentum, was the first to be understood. 
In \cite{Dabholkar:1995nc}, a classical picture of its microstates in terms of an oscillating string was provided. 
This corresponds exactly to the transition \eqref{F1-P_transition} described in the last subsection. 
From the oscillating $\F$-$\W$ configurations, the authors of \cite{Kanitscheider:2007wq} and \cite{Mathur:2005zp} generated fuzzball geometries of the $\D{1}$-$\D{5}$ system by dualities. In the duality map (Fig. \ref{fig:DualityMap}), we can follow this path and see which glues are generated by which internal or external excitations of the string. This path is illustrated at the top of Figure \ref{fig:PathsFromF1PToD1D5}. 
The internal glues of the fundamental string, that is the glues along $T^4_{1234} \times S^1_y$, get mapped to the internal glues of the $\D{1}$-$\D{5}$ system, which are the $\F$-$\NS$ and the $\D{3}$-$\D{3}$ glues. 

It is also possible to take another path from the $\F$-$\W$ system to the $\D{1}$-$\D{5}$ (see the bottom part of Figure \ref{fig:PathsFromF1PToD1D5}). It generates the same glues as the path in \cite{Kanitscheider:2007wq} and \cite{Mathur:2005zp}, but they originate from different oscillations of the fundamental string. The $\F$-$\NS$ glues now comes from the oscillations in the direction $x^4$ instead of $x^1$. The mapping of the external excitations is the same for both paths; the oscillations in $x^9$ become a Kaluza-Klein monopole carrying momentum in the non-compact dimensions.

Let us also remark that not all the glues are necessarily involved in the microscopic degeneracy of brane systems. An obvious example can be found in the $\F[1]$-$\W[1]$ duality frame, which admits $\NS[56789]$-$\KKM[56789]$ as a possible pair of glues (See Figure \ref{fig:P-Dp-Configurations}). But this pair of glues does not contribute to the microscopic degeneracy of the $\F$-$\W$ system, which comes solely from the string's transverse oscillation modes, represented by $\F[2]$-$\W[2]$ glues.\footnote{It would be indeed quite unnatural for this pair of NS5-KKM glues, which involves five additional spatial dimensions to the original $\F[1]$-$\W[1]$, to contribute to its degeneracy. We could perhaps instead view the presence of this pair of glues through the lens of the \textit{inverse supertube transition} (see Sec. \ref{sssec:TheInverseSupertubeTransition}): that an $\NS[56789]$-$\KKM[56789]$ system may have local $\F[1]$-$\W[1]$ polarisations.} 

\subsection{A Remark on Exotic Branes}\label{subsec:RemarkOnExoticBranes}

Exotic Branes naturally appear when dualizing branes. A prime example is the $5_2^2$ brane. It emerges from T dualizing the $\NS$ brane along two perpendicular directions. By using T and S dualities, we can generate many more exotic branes \cite{deBoer:2012ma}. In \cite{Kimura:2016xzd}, the supersymmetric projection rules for exotic branes were derived. The involutions of the presented exotic branes coincide with the involutions of the standard branes. This is summarized in Tables \ref{tab:ExoticBraneInvolutionsForSolitonicBranesIIA}, \ref{tab:ExoticBraneInvolutionsForSolitonicBranesIIB} and \ref{tab:ExoticBraneInvolutionsForDBranes}. 
To derive the duality map and the list of local supersymmetry enhancements, we solely used their respective involutions. Hence, we can interchange any brane (or glue) with an exotic brane that has the same involution, and still have a 1/4-BPS configuration or a local supersymmetry enhancement.

Let us give an example for this. The $\D{1}$-$\D{7}$ system is a non-standard configuration. Using Table \ref{tab:ExoticBraneInvolutionsForDBranes}, we see that the $7_3$ brane, also called the $\mathrm{NS7}$ brane, has the same supersymmetry projector as a $\D{7}$ brane. Therefore, we have found an exotic 1/4-BPS configuration, the $\D{1}$-$\mathrm{NS7}$ intersection, that can be enhanced by including the same glues that would enhance the $\D{1}$-$\D{7}$ system, e.g. 
\be
\[ \, \D[1]{1} \perp \mathrm{NS7}_{2345678} \; (0) \, \] \;
\rightarrow \;
\( \, \NS[12349] \perp \KKM[56789] \, \) \,
\ee
or using exotic branes 
\be
\[ \, \D[1]{1} \perp \mathrm{NS7}_{2345678} \; (0) \, \] \;
\rightarrow \;
\( \, \D[2]{1} \perp \mathrm{NS7}_{1345678} \, \) \, .
\ee

The question arises whether we can connect the standard and non-standard configurations by including exotic branes. Take the S-dual of the standard configuration $\F$-$\D{7}$. This gives the exotic configuration $\D{1}$-$\mathrm{NS7}$ discussed above. If we replace the $\mathrm{NS7}$ by its `normal' counterpart, we arrive at a non-standard configuration, but this is not a duality. We did not find a way to connect the two regimes directly. It seems like including exotic branes does not connect the standard and non-standard brane configurations directly by dualities.

\section{Applications for three-charge systems}
\label{sec:Sec3}

In the previous section, we have classified all local supersymmetry enhancements for 2-charge configurations. Now, we turn to 1/8-BPS systems, \textit{i.e.} 3-charge configurations.
This is especially interesting for black holes, as 3-charge configurations have a finite horizon area at the classical level (without higher-derivative corrections for example). Configurations with 16 local supersymmetries could form a way to describe microstates of these black holes. 

In subsection \ref{subsec:GeneralitiesOfTheThreeChargeSystems}, we will look at the supersymmetry projector of an unspecified 3-charge configuration with six distinct glues. This gives general conditions on the glues and, if they are satisfied, a solution for the charges with two degrees of freedom. We will reduce these requirements to a minimum and provide a procedure to find glues for a given 3-charge system in subsection \ref{subsec:MinimalRequirementsOnA3ChargeLSE}. Finally, we describe an interchange symmetry between the main branes and the glues for the existence of a local supersymmetry enhancement in subsection \ref{subsec:SymmetriesBetweenMainBranesAndGlues}.


\subsection{Generalities of the three-charge systems projector}\label{subsec:GeneralitiesOfTheThreeChargeSystems}

\begin{figure}
    \centering
    \includegraphics[width = 0.4 \textwidth]{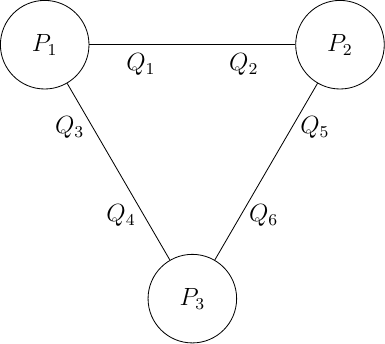}
    \caption{Triangle ansatz of the local supersymmetry enhancement (LSE) for a three-charge, $1/8$-BPS system. The involutions of the main branes are denoted by $P_i$, while the involutions of the glues are denoted by $Q_j$. The glues on the line between two main branes are involved in the two-charge LSE of those two main branes.}
    \label{fig:triangleansatz}
\end{figure}

A natural way to generalize the scheme in the previous section to three charges is to introduce two glues for each pair of main branes.
Denoting the three main branes and their involutions by $P_1$, $P_2$ and $P_3$, this gives six glues with involutions $Q_1$ to $Q_6$. A graphical representation can be found in Figure \ref{fig:triangleansatz}. The projector of what we will call the \textit{triangle ansatz} follows to be
\begin{equation}
    \Pi \equiv \frac{1}{2} \left( 1 + \alpha_1 (x) \, P_{1} + \alpha_2 (x) \, P_{2} + \alpha_3 (x) \, P_{3} 
    + \sum_{j=1}^{6} \beta_j (x) \, Q_{j} \right) \,.
\end{equation}
We assume that the involutions of all main branes and all glues are distinct, i.e. they describe different objects or extend along different directions.\footnote{This excludes the configuration considered in \cite{Bena:2011uw}, see subsection \ref{subsec:D1-D5-P-InternalDipoles}.}

Calculating the projector condition \eqref{eq:projector_condition} and the global supersymmetry condition \eqref{eq:ansatzforglobalsupersymmetry} as for the 2-charge case, there are a number of constraints which also directly follow from imposing a two-charge LSE between each pair of main branes: 
\begin{subequations}\label{eq:unrefined3ChargeEquations}
\begin{align}
    1 &= \alpha_1 + \alpha_2 + \alpha_3 \, , \label{eq:BPSCondition3Charges}\\
    \beta_1 &= - \beta_2 \, \eta_{12} \, , \label{eq:pairOfGlues12}\\
    \beta_3 &= - \beta_4 \, \eta_{13} \, , \label{eq:pairOfGlues34}\\
    \beta_5 &= - \beta_6 \, \eta_{23} \, , \label{eq:pairOfGlues56}\\
    \alpha_1 \alpha_2 &= - \eta_{12} \, \beta_1 \beta_2 \, , \label{eq:quadraticEquation12}\\
    \alpha_1 \alpha_3 &= - \eta_{13} \, \beta_3 \beta_4 \, , \label{eq:quadraticEquation13}\\
    \alpha_2 \alpha_3 &= - \eta_{23} \, \beta_5 \beta_6 \, . \label{eq:quadraticEquation23}
\end{align}
\end{subequations}
Like in \eqref{def_eta}, $\eta_{ij}$ is defined as the sign that appears for the local supersymmetry enhancement between the main branes with involutions $P_i$ and $P_j$,
\begin{subequations}
\begin{align}
    \eta_{12} &\equiv P_1 Q_1 Q_2 P_2 \\
    \eta_{13} &\equiv P_1 Q_3 Q_4 P_3 \\
    \eta_{23} &\equiv P_2 Q_5 Q_6 P_3 \, .
\end{align}
\end{subequations}

In addition to the constraints above, there are 18 new anticommutators appearing in the projector condition \eqref{eq:projector_condition}, \textit{i.e.} in the equation analogous to equation \eqref{eq:2chargeAnticommutatorConstraint}. 
Six of them are between a main brane and a glue on the opposite side of the triangle, the other twelve between two glues on different edges in the diagram. We will cancel some of them with each other by the same procedure as was done for the 2-charge configuration in equation \eqref{eq:cancellationanticommutators}:

Take two general anticommutators with their respective coefficients, $\alpha \, \beta \, \{A,B\}$ and $\gamma \, \delta \, \{C,D\}$.
There are three requirements on cancelling them with each other. First, the product of all involved involutions has to be a sign, $\mu \equiv A B C D = \pm 1$. Second, we have to equate the coefficients up to a sign, $\alpha \beta = - \mu \, \gamma \delta$. And third, we need to have the correct anticommutation relations between the involutions, $\{A,C\}=\{B,D\}=0$. With these three conditions, the calculation is completely analogous to equation \eqref{eq:cancellationanticommutators} for the 2-charge case:

\begin{align}
    \alpha \beta \, \{ A , B \} &=- \gamma \delta \, \mu \, (A B + B A) \nonumber\\
    &=- \gamma \delta \, (A \mu B + B \mu A) \nonumber\\
    &=- \gamma \delta \, (C D + B D B A C A) \nonumber\\
    &=-\gamma \delta \, (C D + D C) = - \gamma \delta \, \{ C , D \} \, . \label{eq:3chargecancellationanticommutators}
\end{align}

In order to have as many degrees of freedom in the coefficients as possible, we will choose the cancellations in a way, such that the corresponding equations are compatible with equations \eqref{eq:quadraticEquation12} to \eqref{eq:quadraticEquation23}. 
\begin{subequations}\label{eq:mixedQuadraticEquations}
\begin{align}
    \alpha_1 \beta_5 = - \mu_{15} \, \beta_2 \beta_3 \label{eq:mu15equation}\\
    \alpha_1 \beta_6 = - \mu_{16} \, \beta_1 \beta_4 \\
    \alpha_2 \beta_3 = - \mu_{23} \, \beta_1 \beta_5 \label{eq:mu23equation}\\
    \alpha_2 \beta_4 = - \mu_{24} \, \beta_2 \beta_6 \\
    \alpha_3 \beta_1 = - \mu_{31} \, \beta_3 \beta_6 \label{eq:mu31equation}\\
    \alpha_3 \beta_2 = - \mu_{32} \, \beta_4 \beta_5 \label{eq:mu32equation}
\end{align}
\end{subequations}

At this point, we introduced six new signs, $\mu_{ij}$. They are defined as the product of the involved involutions and can be read of the equations above, \textit{e.g.} $\mu_{15} = P_1 Q_5 Q_2 Q_3$.\footnote{We use the letter $\eta_{ij}$ to denote a sign involving two main branes $P_i$ and $P_j$ and two glues, while the letter $\mu_{kl}$ uses only one main brane $P_k$ and three glues. The index $l$ is the index of the glue $Q_l$ that commutes with the main brane $P_k$, while the other two involved glues anticommute with $P_k$.} 
These signs have the same properties as $\eta$ and, similarly, fix the (anti-) commutation relations for the involutions. Let us illustrate this for $\mu_{15} = P_1 Q_5 Q_2 Q_3$. $P_1$ and $Q_5$, as well as $Q_2$ and $Q_3$ commute, as otherwise we would not need to cancel their anticommutators in the first place. Furthermore, we already know from the LSE between the main branes $P_1$ and $P_2$, that $P_1$ anticommutes with $Q_2$. Using the invariance under circular of shifts, we can deduce that all other combinations of involutions anticommute by cycling through the involutions one by one. 

Returning to the 18 new anticommutators in the projector condition, the equations \eqref{eq:mu15equation} to \eqref{eq:mu32equation} above allow us to cancel the six anticommutators between main branes and glues on the opposite side of the triangle with six of the other anticommutators. The remaining six anticommutators vanish as a result of the invariance under circular shift of $\mu_{ij}$ as shown above. 

The last step is to find a solution to the equations in \eqref{eq:unrefined3ChargeEquations} and \eqref{eq:mixedQuadraticEquations}. The independent equations are\footnote{In order to derive equations \eqref{eq:quadraticEquation12} to \eqref{eq:quadraticEquation23} from equations \eqref{eq:mixedQuadraticEquations}, one needs the relations \eqref{eq:relationsBetweenEtaAndMu} between the $\eta$s and $\mu$s. They can easily be checked by inserting their definitions and using the derived (anti-)commutation relations.}
\begin{align}
    1 &= \alpha_1 + \alpha_2 + \alpha_3 \, , \tag{\ref{eq:BPSCondition3Charges}}\\
    \beta_1 &= - \beta_2 \, \eta_{12} \, , \tag{\ref{eq:pairOfGlues12}}\\
    \beta_3 &= - \beta_4 \, \eta_{13} \, , \tag{\ref{eq:pairOfGlues34}}\\
    \beta_5 &= - \beta_6 \, \eta_{23} \, , \tag{\ref{eq:pairOfGlues56}}\\
    \alpha_1 \beta_5 &= - \mu_{15} \, \beta_2 \beta_3 \, , \tag{\ref{eq:mu15equation}}\\
    \alpha_2 \beta_3 &= - \mu_{23} \, \beta_1 \beta_5 \, , \tag{\ref{eq:mu23equation}}\\
    \alpha_3 \beta_1 &= - \mu_{31} \, \beta_3 \beta_6 \, . \tag{\ref{eq:mu31equation}}
\end{align}
With the ansatz, $\alpha_1 = a^2$, $\alpha_2 = b^2$ and $\alpha_3 = c^2$, the first equation dictates to choose any parametrization of the 2-sphere for $a$, $b$ and $c$. This can be interpreted as a free choice on how to distribute the energy between the three main branes. Then, the other equations are solved uniquely (up to reorientation of glues) by
\begin{subequations}\label{eq:3ChargeSolution}
\begin{align}
    \beta_1 &= \kappa_1 \, a b \\
    \beta_3 &= \kappa_3 \, a c \\
    \beta_5 &= \kappa_5 \, b c \\
    \beta_2 &= - \eta_{12} \, \beta_1 \\
    \beta_4 &= - \eta_{13} \, \beta_3 \\
    \beta_6 &= - \eta_{23} \, \beta_5 \, ,
\end{align}
\end{subequations}
where $\kappa_i = \pm 1$ dictates the orientation of the glues and satisfies $\kappa_1 \kappa_3 \kappa_5 = - \mu_{23}$.

We can choose any parametrization for the 2-sphere, for example
\begin{equation}
\begin{split}
    a &= \cos \theta_1 = c_1 \\
    b &= \cos \theta_2 \, \sin \theta_1 = c_2 s_1 \\
    c &= \sin \theta_2 \, \sin \theta_1 = s_2 s_1 \, ,
\end{split}
\end{equation}
where $\theta_1 = \theta_1 (x)$ and $\theta_2 = \theta_2 (x)$ are the polar and azimuth angle. 
The supersymmetry projector is then given by
\begin{align}
    \Pi &= \frac{1}{2} \, \big( 1 + c_1^2 \, P_1 + c_2^2 s_1^2 \, P_2 + s_2^2 s_1^2 \, P_3 
    + \kappa_1 \, c_1 c_2 s_1 \, (Q_1 - \eta_{12} \, Q_2) \notag\\
    & \quad + \kappa_3 \, c_1 s_2 s_1 \, (Q_3 - \eta_{13} \, Q_4) 
    + \kappa_1 \kappa_3 \, c_2 s_2 s_1^2 \, (Q_5 - \eta_{23} \, Q_6) \big) \\
    &= c_1 (c_1 + \kappa_1 \, c_2 s_1 \, Q_1 P_1 + \kappa_3 \, s_2 s_1 \, Q_3 P_1) \Pi_1 
    \notag\\ &
    \quad + c_2 s_1 (c_2 s_1 - \kappa_1 \eta_{12} \, c_1 \, Q_2 P_2 + \kappa_1 \kappa_3 \, s_2 s_1 \, Q_5 P_2) \Pi_2 
    \notag\\ &
    \quad + s_2 s_1 (s_2 s_1 - \kappa_3 \eta_{13} \, c_1 \, Q_4 P_3 - \kappa_1 \kappa_3 \eta_{23} \, c_2 s_1 \, Q_6 P_3) \Pi_3 
    \, .
\end{align}
The angles involved in the choice of parameterisation of the two-sphere can sometimes have a geometric interpretation, like in \cite{Li:2023jxb,Bena:2022wpl}.

\subsection{Minimal requirements on a 3-charge LSE}\label{subsec:MinimalRequirementsOnA3ChargeLSE}

In the subsection above, we put a fair amount of assumptions on the involutions of the main branes and glues. Here, we will reduce the requirements to a minimum and provide an easy procedure for finding 3-charge configurations that exhibit local supersymmetry enhancement.

We can derive the following relations between the signs by inserting the definitions. At this point, we can consider all involutions as commuting, because we disregard a total sign.

\begin{subequations}\label{eq:relationsBetweenEtaAndMu}
\begin{align}
    \mu_{15} \mu_{16} &\sim - \eta_{12} \eta_{23} \eta_{13} \\
    \mu_{15} \eta_{12} &\sim - \mu_{23} \\
    \mu_{16} \eta_{12} &\sim - \mu_{24} \\
    \mu_{15} \eta_{13} &\sim - \mu_{32} \\
    \mu_{16} \eta_{13} &\sim - \mu_{31} \, ,
\end{align}
\end{subequations}
where $\sim$ means equality up to a sign.

Therefore, by assuming that $\eta_{12}$, $\eta_{13}$, $\mu_{15}$ and $\mu_{16}$ are signs, we can derive all other signs and also deduce that they are indeed only signs. Furthermore, assuming the correct (anti-)commutation relations in the LSEs for $\eta_{12}$, $\eta_{13}$ and $\eta_{23}$, all others are fixed by the same reasoning as in the subsection above. The correct relations for $\eta_{23}$ can be rewritten into the condition $\{Q_1,Q_3\}=0$ by considering $\mu_{23}$. In this case, the relations \eqref{eq:relationsBetweenEtaAndMu} above hold exactly (and not only up to a sign).

Let us build a 3-charge configuration for three fixed main branes $P_1$, $P_2$ and $P_3$. The first step is to choose glues $Q_1$ to $Q_4$ such that we have two 2-charge LSEs and $\{Q_1,Q_3\}=0$. To this end, we can use the duality maps with all possible LSEs introduced in section \ref{sec:Sec2}, that is Figures \ref{fig:P-Dp-Configurations} to \ref{fig:Dp-NS5-NonStandard-Configurations-2} in the appendix.
Secondly, we have a look at $\mu_{15} \equiv P_1 Q_5 Q_2 Q_3 = \pm 1$. This equation can be satisfied by choosing $Q_5 = \mu_{15} P_1 Q_3 Q_2$.\footnote{A possible sign can be absorbed by the coefficient $\beta_5$. Ultimately, the freedom on choosing the orientation of the branes will be controlled by the signs $\kappa_i$.} This is only possible if $P_1 Q_3 Q_2$ corresponds to an existing brane in the theory. For example, $Q_5 = \Gamma^{0123} \sigma_3$ is not an involution of a brane, as there are no three-dimensional solitonic branes in Type-II string theory. In this case, we can use the exchange symmetry between $Q_1$ and $Q_2$ to get another ansatz for $Q_5$ that might work. The same has to be done for $Q_6$ by utilizing the equation $\mu_{16} \equiv P_1 Q_6 Q_1 Q_4 = \pm 1$. If there are such glues, we have found a 3-charge configuration that exhibits local supersymmetry enhancement. Otherwise, we have to choose different glues $Q_1$ to $Q_4$ and start over.

Let us conclude this part with a remark on the definitions of $\mu_{ij}$, demonstrated on $\mu_{23}$. 
The indices fix the first two involutions of the product defining $\mu_{23}$, $P_2$ and $Q_3$. The other two glues are always on the other two sides of the triangle, that is the sides next to the main brane $P_2$. To determine them in our notation system, we look at the main brane that is next to $Q_3$, in our case $P_1$, and take the other glue that is next to it, $Q_1$. The fourth glue is on the last side of the triangle and right next to the starting main brane $P_2$, in this example $Q_5$. This can be understood best by retracing these steps in Figure \ref{fig:triangleansatz} or the left side of Figure \ref{fig:PermutationSymmetryFor3Charges}
This simple procedure works for all $\mu_{ij}$ and is the only reason for the specific definitions of signs used in this paper. You could easily change this notation by using the symmetry between for example $Q_1$ and $Q_2$ and relabel the glues to end up with a different convention on the signs.

\subsection{A Symmetry between main branes and glues}\label{subsec:SymmetriesBetweenMainBranesAndGlues}

In the 2-charge case, we introduced a symmetry between main branes and glues in subsection \ref{sssec:TheInverseSupertubeTransition}. If there is a local supersymmetry enhancement between the main branes $P_1$ and $P_2$ when including the glues $Q_1$ and $Q_2$, then the configuration with main branes $Q_1$ and $Q_2$ and glues $P_1$ and $P_2$ also exhibits local supersymmetry enhancement.

\begin{figure}[h]
    \centering
    \begin{subfigure}{0.45\textwidth}
        \includegraphics{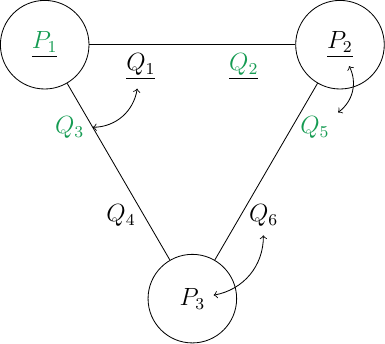}
    \end{subfigure}
    \begin{subfigure}{0.45\textwidth}
        \includegraphics{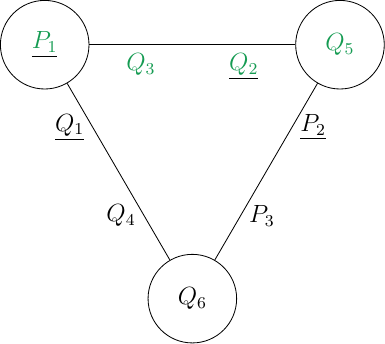}
    \end{subfigure}
    
    \caption{The left figure shows the three-charge configuration before the transformation, the right figure after. The arrows indicate which branes switch roles. 
    The involutions in $\mu_{1,5} = \textcolor{ForestGreen}{P_1 Q_2 Q_3 Q_5}$ are highlighted in green and the involutions in $\eta_{12} = \underline{P_1 Q_1 Q_2 P_2}$ are underlined to demonstrate the switch between $\eta_{12}$ and $\mu_{1,5}$.}
    \label{fig:PermutationSymmetryFor3Charges}
\end{figure}

There is a similar symmetry in the 3-charge case. Let us take an LSE with main branes $P_1$, $P_2$ and $P_3$ and glues $Q_1$ to $Q_6$. The symmetry interchanges $P_2$ with $Q_5$, $P_3$ with $Q_6$ and $Q_1$ with $Q_3$ (see Figure \ref{fig:PermutationSymmetryFor3Charges}). To preserve local supersymmetry enhancement, we need a sign $\eta'_{12}$ for the new configuration which includes both $P_1$ and $Q_5$. This is $\mu_{15}$. Furthermore, the new sign $\mu'_{15}$ now becomes $\eta_{12}$. The same thing happens for $\eta_{13}$ and $\mu_{16}$. We interchange the roles of $\eta$ and $\mu$. This is summarized in Figure \ref{fig:PermutationSymmetryFor3Charges}. 

\begin{figure}[h]
    \centering
    \begin{subfigure}{0.45\textwidth}
        \includegraphics{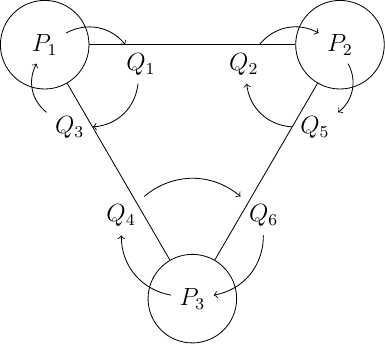}
    \end{subfigure}
    \begin{subfigure}{0.45\textwidth}
        \includegraphics{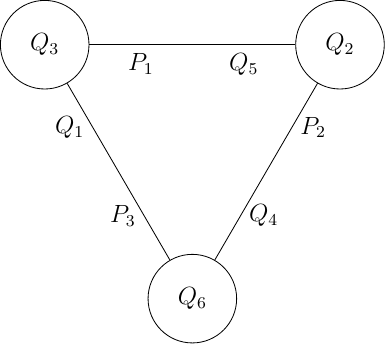}
    \end{subfigure}
    
    \caption{
    The left figure shows the three charge configuration before the transformation, the right figure after. The arrows indicate the permutations. 
    }
    \label{fig:CyclicPermutationSymmetryFor3Charges}
\end{figure}

To permute all main branes with glues simultaneously, we can apply the symmetry twice. Fix $P_1$ and permute $P_2$ with $Q_5$, and $P_3$ with $Q_6$ as before. Afterwards we can fix for example the bottom main brane $Q_6$ and permute $P_1$ with $Q_3$ and $Q_5$ with $Q_2$. The result is depicted in Figure \ref{fig:CyclicPermutationSymmetryFor3Charges}. Both permutations combined result in a clockwise permutation of the main branes with the glues next to them.

\section{Three-charge systems: examples}
\label{sec:Sec3_prime}

In this Section, we discuss four different configurations that fit into the scheme presented in Section \ref{sec:Sec3}:
\begin{itemize}
    \item $\D{1}$-$\D{5}$-$\W$,
    \item $\NS$-$\D{2}$-$\D{6}$,
    \item $\D{0}$-$\F$-$\D{4}$ and 
    \item $\D{0}$-$\D{4}$-$\F$.
\end{itemize}

\subsection{D1-D5-P with internal dipoles}\label{subsec:D1-D5-P-InternalDipoles}

As a first example of supersymmetry enhancement for 3-charge configuration, consider the celebrated $\D{1}$-$\D{5}$-$\W$ system, on $T^4 \times S^1_y$. The circle $S^1_y$ is the usual common D1-D5-P direction, and we label the directions of $T^4$ by $1$,$2$,$3$ and $4$.
The entropy of the black hole is captured, at leading order in the open-string picture, by the number of ways to partition the momentum along $y$ into fractionated momentum quanta carried by an effective D1-D5 string \cite{Strominger:1996sh}. 

Before we begin, our purpose here is not to take such a ``Strominger-Vafa microstate'' (represented by a long D1-D5 string configuration with left-moving momentum) and enhance locally its supersymmetries. As noted in the Introduction, this logic has been applied in \cite{Bena:2022wpl} for Dijkgraaf-Verlinde-Verlinde (DVV) microstates, which backreact into a configuration with maximal local supersymmetry, called the supermaze. This logic worked there because the DVV microstates can be uplifted to a fractionated brane system  -- M2 branes fractionated into M2-strips between pairs of M5-branes (similar to Fig. \ref{fig:NS5-D2_fractionation}) -- that carries integer-mode momenta. In the D1-D5-P frame however, the Strominger-Vafa microstates are essentially one very long effective D1-D5 string that winds $N_1N_5$ times around the $y$ circle; and the entropy comes from the fractionation of the momentum, in units of $1/N_1N_5$. 

Instead, our aim, at this stage, is to identify the ingredients (in terms of local brane charges) that would lead to a local supersymmetry enhancement (LSE) of a brane system with D1, D5 and momentum charges, up to 16 local supercharges. In particular, we would like to ask whether it is possible to enhance the local supersymmetries of the D1-D5-P system with only internal degrees of freedom as glues. 

\begin{figure}[h]
    \centering
    \includegraphics[width = 0.4 \textwidth]{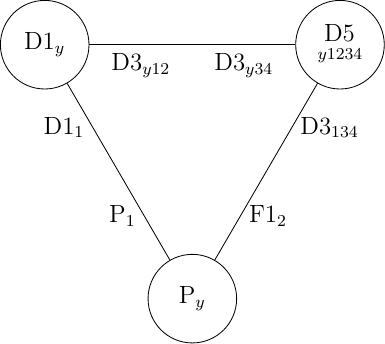}
    \caption{Graphical representation of a local supersymmetry enhancement of the $\D{1}$-$\D{5}$-$\W$ configuration with the respective glues.}
    \label{fig:D1-D5-P}
\end{figure}

And the answer is yes. The option we present here involves the newly discovered D3-D3 glue as a glue in the two-charge D1-D5 subsystem. 
A possible projector associated to this LSE can be written as
\begin{equation}
\begin{split}
    \Pi \equiv \frac{1}{2} ( 1 &+ \alpha_1 \, P_{\D[y]{1}} + \alpha_2 \, P_{\D[y1234]{5}} + \alpha_3 \, P_{\W[y]} \\
    &+ \beta_1 \, Q_{\D[y12]{3}} + \beta_2 \, Q_{\D[y34]{3}} \\
    &+ \beta_3 \, Q_{\D[1]{1}} + \beta_4 \, Q_{\W[1]} \\
    &+ \beta_5 \, Q_{\D[134]{3}} + \beta_6 \, Q_{\F[2]} ) \, .
\end{split}
\end{equation}
This LSE corresponds to a triangular configuration and is illustrated in Figure \ref{fig:D1-D5-P}.

As stated before, we are involving the orthogonal D3-D3 glues between the D1-D5 two-charge subsystem.
Concerning the two-charge $\D{1}$-$\W$ subsystem along $y$, the most natural LSE is to give the D1 a shape within $T^4$, and make it independent of one of the light-like coordinates of $y$. This gives the $\D[1]{1}$-$\W[1]$ glues, where the coordinate $x^1$ labels the extension of the D1 profile inside the $T^4$. 
Lastly, a consistent way for a D5 brane to carry momentum along $y$ that involves only internal excitations is to have positive and negative densities of orthogonal F1-D3 glues. This is very similar to the way the NS5 branes can carry momentum along $y$ using D0 and $\D[1234]{4}$ dipole charges as in \cite{Bena:2022wpl,Bena:2022sge}. Note that in order to have a consistent LSE, the orthogonal F1-D3 glue cannot be a $\F[1]$-$\D[234]{3}$ glue, as the direction $x^1$ has been already used for the D1-P glue. Given the directions chosen for the glues between D1-D5 and D1-P, the only possibility between D5-P is to have the directions $\F[2]$-$\D[134]{3}$. 
Note also that this LSE of the D1-D5-P system with only internal excitations is S- and T-dual to the LSE of the F1-NS5-P system with internal excitations of \cite{Bena:2022wpl} (the projector of the super-maze). 

As a matter of fact, it was believed to be impossible to maximally enhance the local supersymmetries of the D1-D5-P system with only internal glues \cite{Bena:2014qxa,Bena:2011uw}. The reason for this is that in \cite{Bena:2011uw}, the authors succeeded in enhancing the D1-D5-P system up to 16 local supersymmetries, but they had to involve degrees of freedom on the $\mathbb{R}^4$ orthogonal to the $S^1_y\times T^4$. Their method consisted in making a two-step supertube transition. First, the $\D{1}$-$\W$ and the $\D{5}$-$\W$ systems undergo separately a supertube transition, by taking both the same spiral shape in $S^1_y\times \mathbb{R}^4$. Second, in order to glue the D1-P supertube with the D5-P supertube, one puffs out the D1-D5 spiral consistently with a $\KKM$ with momentum, which extends along a closed curve in $S^1_y\times \mathbb{R}^4$, perpendicular to the D1-D5 spiral. The glues used in this two-step supertube transition exhibit some symmetry. However this symmetry is not that of the Triangle ansatz of Fig. \ref{fig:triangleansatz}. This is how our LSE in Fig. \ref{fig:D1-D5-P} can avoid the no-go, so that it involves only internal degrees of freedom.

\subsection{NS5-D2-D6: combining two ways to generate entropy} \label{subsec:NS5-D2-D6}

In this subsection, we study the $\NS$-$\D{2}$-$\D{6}$ configuration inside $S^1_y \times S^1_z \times T^4_{1234}$. We label the three remaining $\mathbb{R}^3$ directions by $x^7$, $x^8$ and $x^9$. The $\D{6}$ extends along the whole torus and both circles, while the $\D{2}$ only extends along the circles but is smeared along $T^4_{1234}$. The $\NS$ extends along the torus and the first circle $S^1_y$ and is smeared along the second one $S^1_z$. This is summarized in Table \ref{tab:NS5-D2-D6-Smearings}. 
\begin{table}[h]
    \centering
    \begin{tabular}{c c c c c c c c c c c}
                & $0$  & $y$  & $z$    & $1$    & $2$    & $3$    & $4$    & $7$     &  $8$    & $9$     \\
        $\D{2}$ & $\--$ & $\--$ & $\--$   & $\sim$ & $\sim$ & $\sim$ & $\sim$ & $\cdot$ & $\cdot$ & $\cdot$ \\
        $\D{6}$ & $\--$ & $\--$ & $\--$   & $\--$   & $\--$   & $\--$   & $\--$   & $\cdot$ & $\cdot$ & $\cdot$ \\
        $\NS$   & $\--$ & $\--$ & $\sim$ & $\--$   & $\--$   & $\--$   & $\--$   & $\cdot$ & $\cdot$ & $\cdot$ 
    \end{tabular}
    \caption{Illustration of where the branes are extended along ($\--$), smeared along ($\sim$) or are pointlike ($\cdot$)}
    \label{tab:NS5-D2-D6-Smearings}
\end{table}
If we considered also a momentum charge P along $y$, then the resulting brane system would give rise to a finite-size black hole in four dimensions. But without the momentum, the three brane species NS5, D2 and D6 all wrap $S^1_y$, thus shrinking its size at the locus of the branes. As such, the resulting four-dimensional supergravity solution is a black hole with a horizon of vanishing area.

\begin{figure}[htb]
    \centering
    \includegraphics[scale=1]{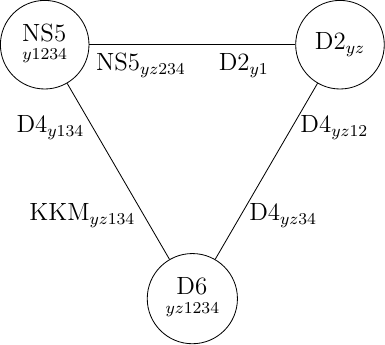}
    \caption{Graphical representation of a local supersymmetry enhancement of the $\NS$-$\D{2}$-$\D{6}$ configuration with the respective glues. Be aware that the glues between the $\NS$ and $\D{6}$ on the left side of the triangle are arranged according to their interpretation in M-theory; to be consistent with the conventions in Section \ref{sec:Sec3}, one needs to switch the place of those two glues in the triangle diagram.}
    \label{fig:NS5-D2-D6}
\end{figure}

It is possible to enhance the supersymmetries of the three-charge system up to 16 supercharges. The configuration with the added glues is illustrated in Figure \ref{fig:NS5-D2-D6}. The resulting supersymmetry projector is
\begin{equation}
\begin{split}
    \Pi \equiv \frac{1}{2} ( 1 &+ \alpha_1 \, P_{\NS[y1234]} + \alpha_2 \, P_{\D[yz]{2}} + \alpha_3 \, P_{\D[yz1234]{6}} \\
    &+ \beta_1 \, Q_{\NS[yz234]} + \beta_2 \, Q_{\D[y1]{2}} \\
    &+ \beta_3 \, Q_{\KKM[yz134]} + \beta_4 \, Q_{\D[y134]{4}} \\
    &+ \beta_5 \, Q_{\D[yz12]{4}} + \beta_6 \, Q_{\D[yz34]{4}} ) \, .
\end{split}
\end{equation}

\begin{figure}[h]
    \centering
    \includegraphics[width=0.7\textwidth]{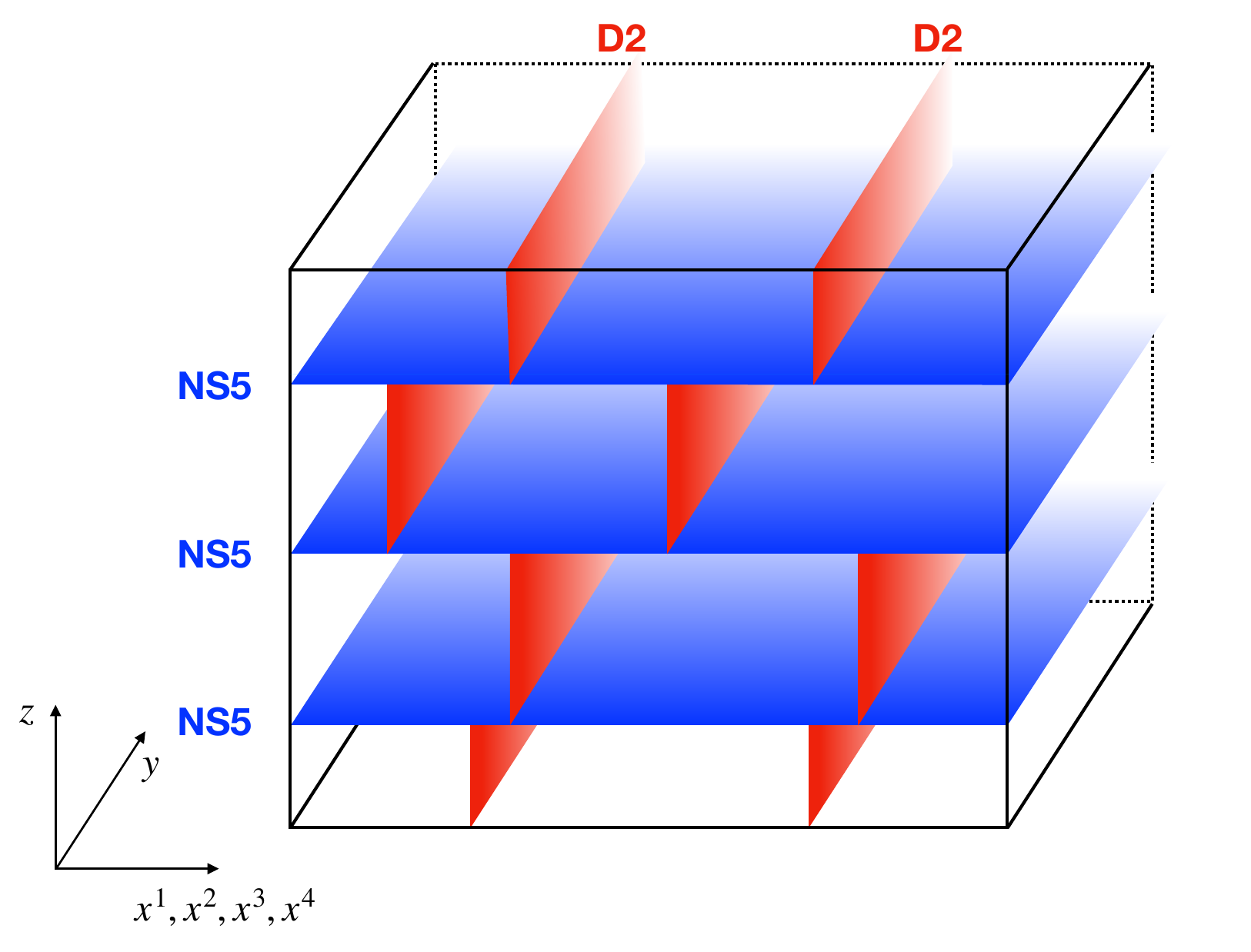}
    \caption{Fractionation of D2 branes between NS5 branes.}
    \label{fig:NS5-D2_fractionation}
\end{figure}

At zero string coupling, the $\NS$-$\D{2}$-$\D{6}$ and $\NS$-$\D{2}$-$\D{6}$-$\W$ configurations combine two different types of entropy generating fractionation. First, without the presence of the NS5 branes, the D2 branes are purely inside the D6 branes, and in particular share the common direction $S^1_y$. Thus, this is dual to the D1-D5 system, and the open-string theory describes an effective long D2-D6 string of length $N_2N_6$. Besides, in the presence of the NS5 branes, the $N_2$ D2 branes fractionate into $N_2N_\mathrm{NS5}$ D2 strips between the NS5 branes, as illustrated in Figure \ref{fig:NS5-D2_fractionation}. As such, the microscopic entropy, that is proportional to $\sqrt{N_2N_\mathrm{NS5}N_6}$, results from the combination of the two above-mentioned fractionation effects \cite{Maldacena:1996gb} -- one may think of it as coming from an effective D2-NS5-D6 string of length $N_2N_\mathrm{NS5}N_6$. 

When one tunes the string coupling to slightly larger values, the branes start interacting. In particular, the D2 branes pull on the worldvolume of the NS5 branes, forming NS5-D2 furrows. This is the meaning of the NS5-D2 glue between the main NS5 and D2 charges (top side of the triangle in Fig.~\ref{fig:NS5-D2-D6}). 
On the right side of the triangle in Fig. \ref{fig:NS5-D2-D6}, we would like to encode the fact that the D2 brane can consistently dilute inside the D6 brane; thus, we use the orthogonal D4-D4 glue between them.

Finally, the only consistent way to complete the triangle diagram with glues on the left of Fig.~\ref{fig:NS5-D2-D6} is to use the D4-KKM pair. Given the symmetries of the system, the special direction of the KKM should be $x^2$, orthogonal to the D4 glue. In M theory, this LSE can be understood as
\be \label{KKM-M5_transition}
\[ \, \mathrm{M5}_{y1234} \parallel \mathrm{KKM}_{yz1234;11} \; (5) \, \] \;
\rightarrow \;
\( \, \mathrm{M5}_{y134\,11} , \mathrm{KKM}_{yz134\,11;2} \, \) \, .
\ee
Without the presence of the glues, as one approaches the location of the branes in $\mathbb{R}^3$, say $(x^7,x^8,x^9)=(0,0,0)$, the M-theory circle (parameterised by $x^{11}$) shrinks smoothly to a point.
The addition of the $( \, \M[y134\,11]{5} , \KKM[yz134\,11;2] \, )$ glue in \eqref{KKM-M5_transition} has the effect of modifying the original M5-KKM system in the $(x^2,x^{11})$ plane. Indeed, the projector of the bound state indicates that we are dealing locally with a pair of orthogonal KKM-M5 branes with 16 supersymmetries, in the directions
\be \label{KKM_perp_to_M5}
\mathrm{KKM}_{yz1\hat{2}34;\hat{11}} \perp \mathrm{M5}_{y1 \,\hat{11}\, 34} \, ,
\ee
where the coordinates $(\hat{x}^2, \hat{x}^{11})$ are a mixture of $(x^2, x^{11})$. This is similar to the F1-P transition \eqref{F1-P_transition}, where the $\mathrm{F1}_y \parallel \mathrm{P}_y$ system in the presence of the glues becomes $\mathrm{F1}_{\hat{y}} \perp \mathrm{P}_{\hat{\psi}}$, describing a F1 string along $\hat{y}$ boosted along its orthogonal direction, $\hat{\psi}$.

One may not take \eqref{KKM_perp_to_M5} at face value, and interpret $\hat{x}^{11}$ as a normal KKM special direction which shrinks smoothly to a point as one approaches $(x^7,x^8,x^9)=(0,0,0)$, since there is a M5-brane charge along $\hat{x}^{11}$. 
But the point we are making here is that the KKM-M5 system \eqref{KKM_perp_to_M5} extends along the $(\hat{x}^2, \hat{x}^{11})$ directions, which make an angle, $\alpha$, with $(x^2, x^{11})$. And the brane densities in the Type IIA system depend on this angle. Thus, making the angle vary in space in the M-theory picture means that in Type IIA, one makes the brane densities depend on the spatial coordinates. 
The quantity of the local charge of the three pairs of glue can be regarded as a response to the quantity of local densities of the main branes as a function of the internal coordinates, following the equations \eqref{eq:3ChargeSolution}.

\subsection{D0-F1-D4 and supertubes ending on branes}
\label{subsec:D0-F1SupertubeEndingOnD4Branes}

\begin{figure}[htb]
    \centering
    \begin{subfigure}{0.4\textwidth}
        \includegraphics{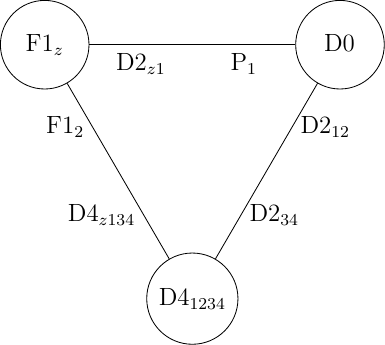}
        \caption{Possible LSE with D2-D2 glues between D0 and D4}
        \label{fig:first}
    \end{subfigure}
    \hfill
    \begin{subfigure}{0.4\textwidth}
        \includegraphics{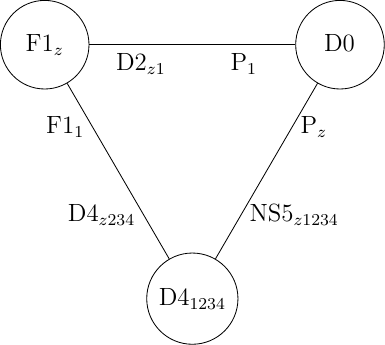}
        \caption{Possible LSE with NS5-P glues between D0 and D4}
        \label{fig:second}
    \end{subfigure}
    \caption{Graphical representations of two possible local supersymmetry enhancements (LSE) of the D0-F1-D4 configuration with the respective glues.}
    \label{fig:F1-D0-D4-possibilities}
\end{figure}

As the third example of a three-charge configuration, we examine the D0-F1-D4 system. Consider Type IIA theory on $T^4\times S^1_z$, where the $T^4$ is labeled by the directions $(1,2,3,4)$. As represented by Figure \ref{fig:F1-D0-D4-possibilities}, the three charges that define the black-hole solution consist of D0 branes smeared along $T^4\times S^1_z$, F1 strings wrapped along $z$ and smeared along $T^4$, and D4 branes wrapped along $T^4$ and smeared on $S^1_z$. After dimensional reduction, it is a five-dimensional black-hole solution with a finite horizon area. We would like to find a local supersymmetry enhancement (LSE) of this three-charge system with all the glues along the internal directions.

An interesting feature about this system is that its microstructure combines two supersymmetry enhancement effects; as such, the overall degeneracy of the three-charge brane system would result from the combination of two types of two-charge brane degeneracies. Namely, this 3-charge configuration combines the effects of a D0-F1 supertube transition with those of a F1-D4 Callan-Maldacena transition. On the one hand, the interaction between a fundamental string $\F[z]$ with a constant $\D{0}$ charge combines into a D2-P supertube of arbitrary shape, see Section \ref{subsec:DiscussionOfTheLocalSupersymmetryEnhancements} and eq. \eqref{eq:supertube_transition_F1D0}. In terms of coordinates, we choose to form the supertube using the $T^4$ directions , say a $\D[z1]{2}$ brane stabilized by a momentum $\W[1]$. 
On the other hand, the interaction between a fundamental string $\F[z]$ and $\D[1234]{4}$ brane results in the formation of a D4-F1 Callan-Maldacena (or BIon) spike.

In a nutshell, one expects the microstructure in the D0-F1-D4 frame to combine the supertube and BIon spike effects, but it is not known in the literature if this combination is possible. We argue that if it is possible, then the resulting microstructure should have 16 local supersymmetries. And within the triangle ansatz, this means that we must fix the type of glues between the F1 and the D0 main branes (top side of the triangles in the Figures of \ref{fig:F1-D0-D4-possibilities}), as well as that between the F1 and D4 main branes (left side of the triangles in the Figures of \ref{fig:F1-D0-D4-possibilities}).

There are two consistent possibilities to complete the glues on the right side of the triangle, in Figure \ref{fig:F1-D0-D4-possibilities}. The difference between the two lies in whether the `radial' direction of the Callan-Maldacena (CM) spikes is the direction of the supertube profile (Figure \ref{fig:second}) or not (Figure \ref{fig:first}). Note that the possible glues on the right side of the triangle are not of the same nature.

Understanding how the microstructure can be geometrically realised in both possibilities in Fig. \ref{fig:F1-D0-D4-possibilities} is beyond the scope of this paper. However, we may have an interpretation. At zero string coupling, one deals with fractionated F1-D0 rods between D4 branes. When one starts increasing the coupling, the F1 pulls the worldvolume of the D4, but at the same time, it induces a D2-P supertube with the D0 charge. For the LSE of Fig. \ref{fig:first}, the radial direction of the CM spike ($x^2$) is orthogonal to the supertube profile (along $x^1$). So instead of a F1 rod pulling the worldvolume of the D4 in the shape of a spike, one has rather a D2-P supertube (with F1 charge) pulling on the D4 -- the resulting D4 worldvolume makes a hole of the shape of the D2-P supertube in the region between two layers of D4 branes (in the $z$ axis). Then, the D0 charge of the supertube dissolves into the D4 worldvolume with a D2-D2 dipole charge around it.

\begin{figure}[h]
    \centering
    \includegraphics[scale=1]{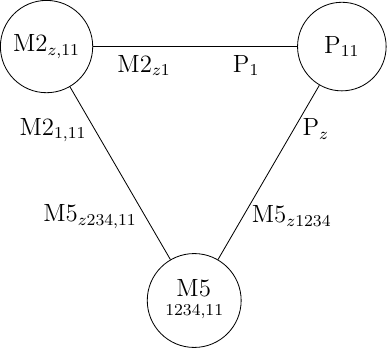}
    \caption{M-Theory uplift of the $\F$-$\D{0}$-$\D{4}$ configuration in Fig. \ref{fig:second}. The M-theory direction is $x^{11}$.}
    \label{fig:MTheoryUpliftOfF1D0D4}
\end{figure}

But for the LSE of Fig. \ref{fig:second}, the radial direction of the CM spike ($x^1$) is parallel to the would-be supertube profile (along $x^1$). Thus, the D2-P glue between the F1 and D0 main branes may not be interpreted as the emergence of a D2-P supertube along a closed profile, along $x^1$. Instead, a better way to understand the LSE is in M theory, see Figure \ref{fig:MTheoryUpliftOfF1D0D4}. 

In Fig. \ref{fig:MTheoryUpliftOfF1D0D4}, we recognise the brane composition of a `supermaze' \cite{Bena:2022wpl}. Without the momentum charge, $\mathrm{P}_{11}$, the projector associated to the diagram describes the backreaction of fractionated M2 strips inside M5 branes. The result of the backreaction is similar to a collection of F1-D4 Callan-Maldacena spikes, but has a common direction, $x^{11}$, so it is rather a collection of M2-M5 furrows. This collection of furrows has the shape of a maze in the $(x^{11},x^1)$ plane, hence the name ``supermaze''. Putting the momentum charge $\mathrm{P}_{11}$ back, the entire triangle diagram of Fig. \ref{fig:MTheoryUpliftOfF1D0D4} describes how this supermaze carries momentum consistently along $x^{11}$.
Performing the dimensional reduction along $x^{11}$, one recovers the LSE diagram of Figure \ref{fig:second}. Therefore, the geometric interpretation of this LSE is about the way for a F1-D4 supermaze to consistently carry a D0 charge along its worldvolume.

\begin{figure}[h]
    \centering
    \begin{subfigure}{0.4\textwidth}
        \includegraphics{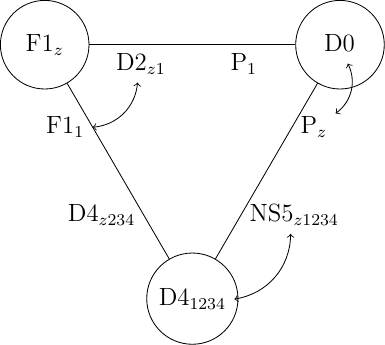}
        \caption{Action of the main-brane/glues symmetry on Fig. \ref{fig:second}} 
        \label{subfig:F1-D0-D4}
    \end{subfigure}
    \hfill
    \begin{subfigure}{0.4\textwidth}
        \includegraphics{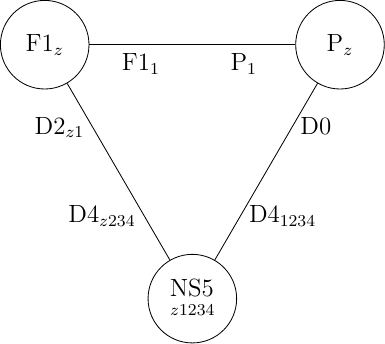}
        \caption{The Type-IIA `supermaze' configuration}
        \label{subfig:supermaze}
    \end{subfigure}
    
    \caption{Using the symmetry between main branes and glues (see Figure \ref{fig:PermutationSymmetryFor3Charges}), illustrated in Fig. \ref{subfig:F1-D0-D4} by the arrows, we can generate the Type-IIA supermaze diagram \ref{subfig:supermaze}.} 
    \label{fig:SymmetryBetweenF1D0D4AndSupermaze}
\end{figure}

Finally, let us notice that one can perform the dimensional reduction along the M2 direction, $z$, and then relabel the direction $x^{11}$ by $z$: One gets the diagram in Fig. \ref{subfig:supermaze}. This describes the LSE of an NS5-F1-P system along $z$, see also \cite{Bena:2022wpl}. It has a close relationship with the F1-D0-D4 system of \ref{fig:second}. Namely, Fig. \ref{subfig:supermaze} can be engineered from \ref{fig:second} thanks to the main-branes/dipoles symmetry of Figure \ref{fig:PermutationSymmetryFor3Charges}, see Fig. \ref{subfig:F1-D0-D4}. This is a nice illustration of the discussion in subsection \ref{subsec:SymmetriesBetweenMainBranesAndGlues}.

\subsection{D0-D8-F1 and the Hanany-Witten effect}
\label{ssec:D0-D8-F1_Hanany-Witten}

We illustrate now the possibility to have a three-charge LSE involving two 1/4-BPS standard brane intersections and one 1/4-BPS non-standard one, namely the D0-D8 intersection. The previous examples involved only standard brane intersections.

\begin{figure}[htb]
    \centering
    \begin{subfigure}{0.4\textwidth}
    \includegraphics{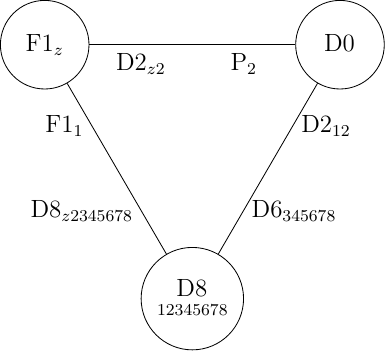}
    \caption{LSE with D2-P glue between F1-D0.}
    \label{fig:D0-D8-F1_with_D2-P}
    \end{subfigure}
    \hfill
    \begin{subfigure}{0.4\textwidth}
    \includegraphics{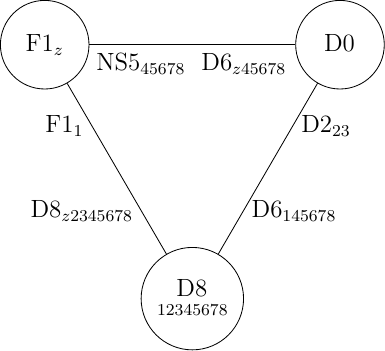}
    \caption{LSE with NS5-D6 glue between F1-D0.}
    \label{fig:D0-D8-F1_with_D6-NS5}
    \end{subfigure}
    \caption{Graphical representation of a local supersymmetry enhancement of the $\NS$-$\D{2}$-$\D{6}$ configuration with the respective glues.}
    \label{fig:F1-D0-D8-possibilities}
\end{figure}

The system we consider is a D0-D8-F1 system, with the F1, along $z$, perpendicular to the D8, spanning along the directions 12345678.
Figures \ref{fig:D0-D8-F1_with_D2-P} and \ref{fig:D0-D8-F1_with_D6-NS5} represent two possible three-charge LSE. 

\begin{figure}[htb]
    \centering
    \begin{subfigure}{0.45\textwidth}
    \begin{adjustbox}{max totalsize={\textwidth}{\textheight},center}
    \includegraphics{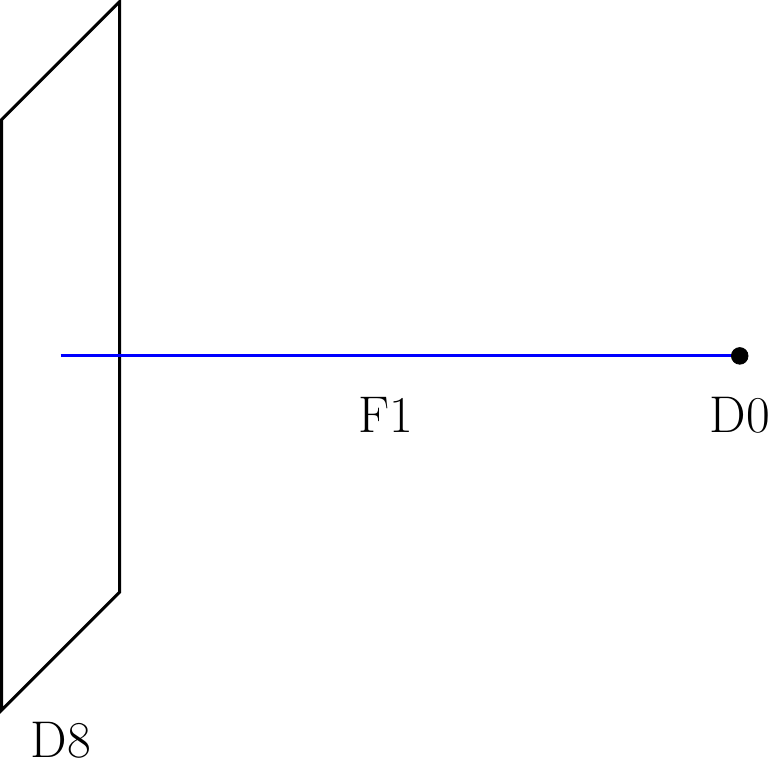}
    \end{adjustbox}
    \caption{The Hanany-Witten effect.}
    \label{fig:D0-D8-F1_Hanany-Witten}
    \end{subfigure}
    \hfill
    \begin{subfigure}{0.45\textwidth}
    \begin{adjustbox}{max totalsize={\textwidth}{\textheight},center}
    \includegraphics{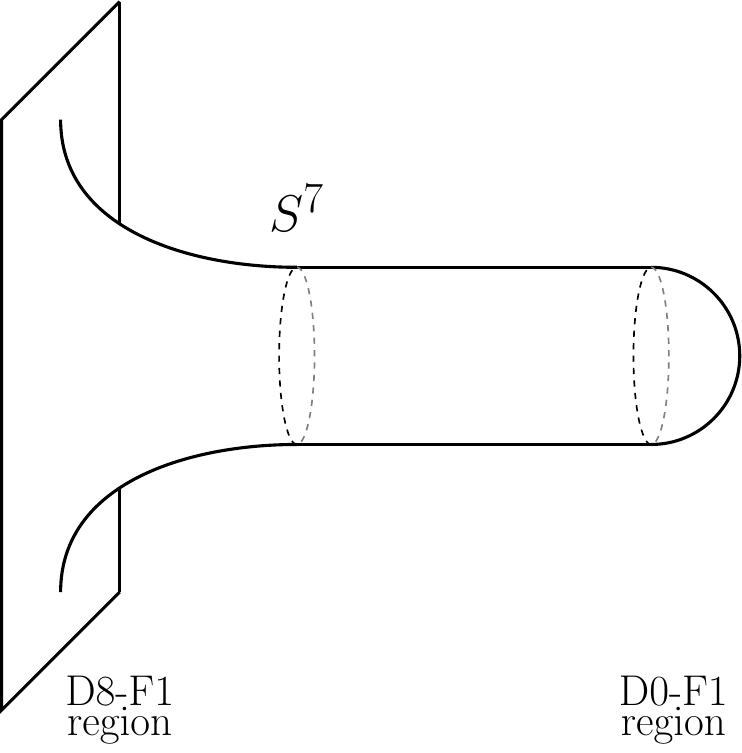}
    \end{adjustbox}
    \caption{Its zoom-in version.}
    \label{fig:D0-D8-F1_microscopic}
    \end{subfigure}
    \caption{The Hanany-Witten effect for the D0-D8 system at zero string coupling (left), and a schematic depiction of its backreaction when $g_s$ increases and the branes start interacting (right).}
    \label{fig:Hanany-Witten_Geometry}
\end{figure}

The possibility of a LSE for the D0-D8-F1 main branes suggests the existence of a local or zoom-in description of the Hanany-Witten effect, in the same way that the LSE of the F1-D3 main branes reveals a local description of a fundamental string ending on a D3 brane in terms of a Callan-Maldacena spike, when the string coupling, $g_s$, is weak but non vanishing.
The Hanany-Witten effect \cite{Hanany:1996ie} for the D0-D8 system \cite{Polchinski:1995sm,Bergman:1997gf} is the fact that moving a D0 brane across a D8 brane creates a F1 string stretching between the two, see Figure \ref{fig:D0-D8-F1_Hanany-Witten}. (Or equivalently, moving a D0 brane away from a D8 brane in either sides creates a F1 string of half-unit charge.) 
The resulting F1 string has a net charge (and not a dipole charge), so it should be considered as a main brane. Thus, the D0-D8 Hanany-Witten effect would involve three main branes. 


Only one of the two possible LSEs in Fig. \ref{fig:F1-D0-D8-possibilities} (at most) can describe the physics of the D0-D8-F1 Hanany-Witten effect. 
In either of the two cases, we expect the resulting 16-local-supersymmetries bound state, which has the topology of an eight-dimensional plane, to take the shape of Fig. \ref{fig:D0-D8-F1_microscopic} when the branes start interacting. Each value of $z$ makes a slice of a seven-dimensional manifold, $X_7$, parameterised by the coordinates $(x^2, \dots, x^8)$ and with the topology of a seven-sphere, $S^7$. Close to the D8-F1 transition region, the F1 pulls into the D8, forming a D8-F1 Callan-Maldacena-type eight-dimensional spike \eqref{Dp-F1_with_Dp-F1}. Without entering into the details, the LSE of Fig. \ref{fig:D0-D8-F1_with_D2-P} seems incompatible with the shape of Fig. \ref{fig:D0-D8-F1_microscopic}. One expects indeed to have the maximum of the D0 charge at the right extremity of Fig. \ref{fig:D0-D8-F1_microscopic}, but the properties of the D2-P supertube would mean that the size of $X_7$ would have to blow up in that region.


Fig. \ref{fig:D0-D8-F1_microscopic} is reminiscent of microscopic descriptions of the Hanany-Witten effect in the literature \cite{Pelc:2000kb,Marolf:2001se}, in other duality frames. It will be interesting to check whether such description of the Hanany-Witten effect in the D0-D8 frame would exhibit the local brane densities of Fig. \ref{fig:D0-D8-F1_with_D6-NS5}. We will leave it for a future endeavour.

\section{Summary and Conclusion}
\label{sec:Concl}

The main focus of this paper is about the concept of local supersymmetries. In a given duality frame (where one measures non-vanishing global charges of some brane species), local supersymmetries reveal the geometric shapes and brane densities of microstructures that can be realised in that duality frame. This notion tells us about how branes of different species can interact and backreact on one another,\footnote{By backreaction, we mean for example that a F1 string ending on a D3 brane backreacts into a D3-F1 Callan-Maldacena spike.} and is particularly relevant when one looks for the local brane composition of pure microstates of BPS black holes. 

First, in Section \ref{sec:Sec2}, we performed a comprehensive characterisation of all possible local supersymmetries in all Type II string-theory $1/4$-BPS backgrounds. The $1/4$-BPS backgrounds consist of all possible brane intersections -- branes within branes, branes intersecting branes, and branes ending on branes -- and are not only relevant in the study of BPS black holes (`standard' brane intersections), but also of BPS domain walls (`non-standard' brane intersections). 

The way to quickly read our numerous tables characterising the possible dipole charges (or \textit{glues}) allowed in every $1/4$-BPS brane-intersection background is the following. Step one, find the desired $1/4$-BPS system in the two `duality maps': Figure \ref{fig:DualityMap} for `standard' brane intersections and Figure \ref{fig:DualityMapNonStandard} for `non-standard' brane intersections. Step two, go to Appendix~\ref{sec:appendix_LSE_tables}, and all the possible glues are listed below the name of each $1/4$-BPS backgrounds.


For a number of possibilities for local supersymmetry enhancement (LSE), we know a physical realisation. They are summarised in Section \ref{subsec:DiscussionOfTheLocalSupersymmetryEnhancements}. 

Then, in Section \ref{sec:Sec3} and \ref{sec:Sec3_prime}, we used these possible local supersymmetry enhancements of $1/4$-BPS systems as the building blocks for studying three-charge, $1/8$-BPS configurations. In particular, we derived in Section \ref{sec:Sec3} necessary and sufficient conditions for an LSE to be realised for $1/8$-BPS systems, within their most natural Ansatz, the `Triangle Ansatz'. 

In Section \ref{sec:Sec3_prime}, this technology enabled us to reveal the possible local brane composition of the microstructure of pure black-hole microstates of $1/8$-BPS black holes. For this, we studied the examples of D1-D5-P, NS5-D2-D6 and D0-F1-D4 systems, for which we locally enhanced the supersymmetries up to 16 supercharges.
Our results suggest that pure microstates in these three systems can carry internal microstructure. Besides, local supersymmetry enhancement is compatible with combining different brane-interaction effects which generate the black-hole entropy -- especially in the NS5-D2-D6 and D0-F1-D4 systems. Furthermore, on the non-standard brane intersection side, we found a three-charge local supersymmetry enhancement that may describe the zoom-in picture of the Hanany-Witten effect for the D0-D8 system.

Let us end this paper with some thoughts about the distinction between internal and external microstructures.
Following the discussion in Section \ref{subsec:dofs_ontopof_background}, in all 1/4-BPS frames, there are glues corresponding to internal and external degrees of freedoms. A natural question is whether the two-charge and three-charge black-hole entropy comes from only internal degrees of freedom, from only external degrees of freedom, or from both.

With some exceptions \cite{Kanitscheider:2007wq,Bena:2022wpl}, the bulk of the effort in the Fuzzball programme has mainly focused on external microstructures, coming from the puffing out of the D1-D5 system with a KKM-P dipole (\textit{e.g.} \cite{Lunin:2001fv,Bena:2011uw,Bena:2017xbt}). But recent works indicate that the locally $1/2$-BPS states with external microstructures should not be thought as the microstates of the black hole, but rather ``BPS stars'' \cite{Martinec:2023xvf,Martinec:2023gte}, because their typical extension is parametrically larger than the horizon scale. 
This would mean, in our language, that the `microstates' which involve KKM-P-type degrees of freedom are not the typical microstates of the D1-D5 black hole.

This work provides a language, in terms of glues, for describing external and internal microstructure. We hope that this language will clarify what degrees of freedom contribute to the black-hole entropy, and what degrees of freedom constitute typical black-hole microstates.

Finally, we would like to point out that our results may predict the existence of a new type of (internal) microstructure, namely the D2-D2 glues inside the D0-D4 system. 
If this D2-D2 glue really exists as a dipolar degree of freedom on top of a D0-D4 background, then these degrees of freedom should be mapped to those of the D0-D4 instanton (see \textit{e.g.} \cite{Tong:2005un,Belitsky:2000ws,Dorey:2002ik,Taylor:1997dy} for reviews). 
In fact, this very type of microstructure is responsible for the LSE in the D1-D5-P frame, by taking the avatar of a D3-D3 glue between the D1 and D5 main brane charges. 
It is unclear why the black-hole entropy in the D1-D5-P frame \cite{Sen:1995in,Strominger:1996sh} comes from momentum-type fractionation, while the entropy in the F1-NS5-P \cite{Dijkgraaf:1996cv} frame or in the M5-M5-M5-P \cite{Maldacena:1997de} frame is due to the interaction of orthogonal branes. 
Perhaps these D3-D3 glues are dipoles which carry degrees of freedom corresponding to fractionated D1 charges inside the D5 worldvolume, and if so, this would be an important piece of data for answering the above puzzle.

\vskip 20pt
	\noindent {\bf Acknowledgements:} We would like to thank Iosif Bena, Emil Martinec, Samir Mathur and Masaki Shigemori for useful discussions. The work of YL is supported by the German Research Foundation through a German-Israeli Project Cooperation (DIP) grant "Holography and the Swampland". 

\appendix

\section{The building blocks along the Duality Map}
\label{sec:appendix_LSE_tables}

\subsection{P - Dp and F1 - Dq configurations}


\begin{figure}[h!]
    \centering
    \includegraphics[width = \textwidth]{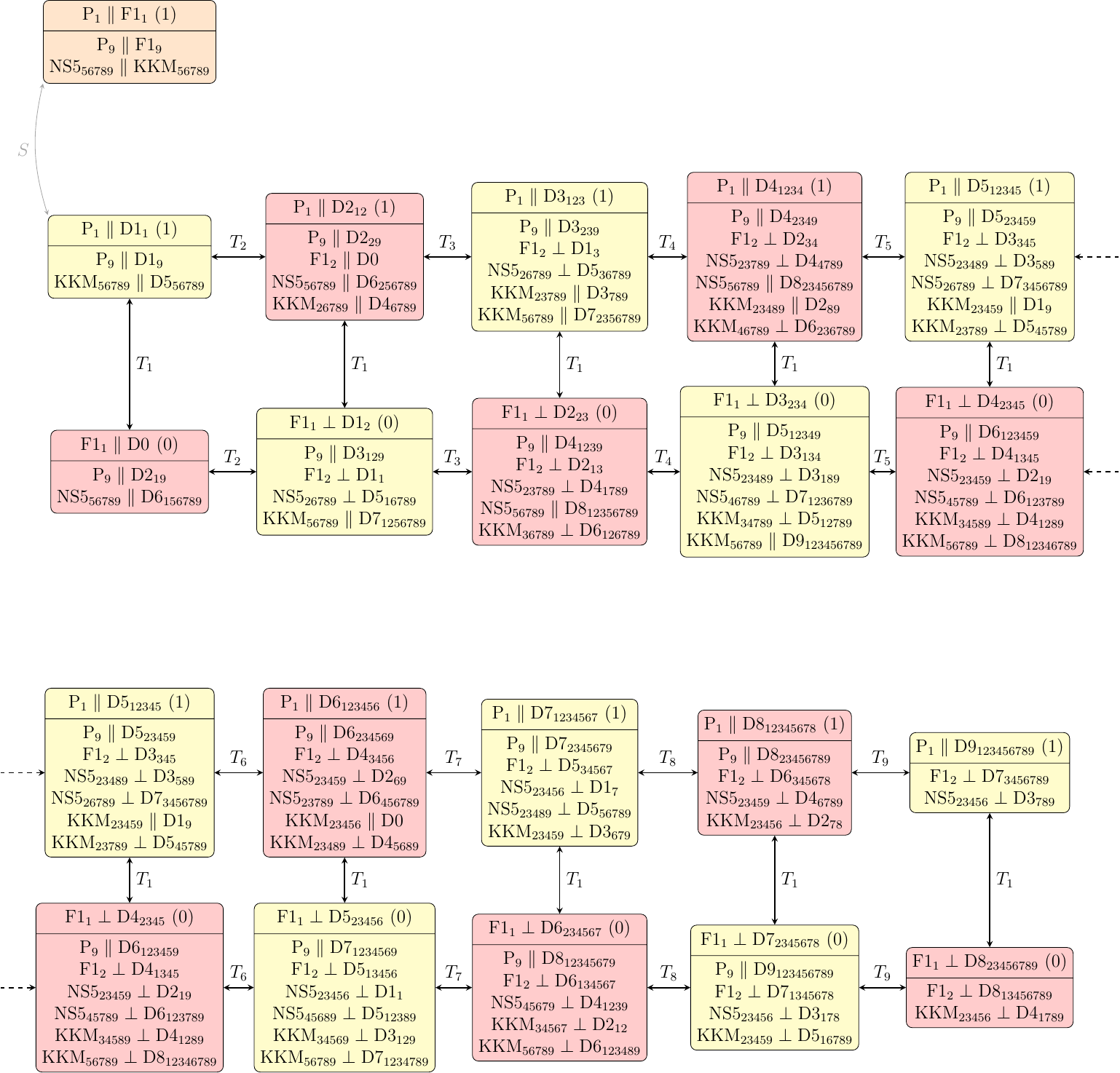}
    \caption{Possible local supersymmetry enhancements of $\W$-$\D{p}$ and $\F$-$\D{q}$ standard configurations. For details on colouring and notation, we refer to the text, especially the segment on notation \ref{text:notation}.}
    \label{fig:P-Dp-Configurations}
\end{figure}

\subsection{Dp - Dq configurations}
\begin{sidewaysfigure}[h!]\footnotesize
    \centering
    \includegraphics[width = 0.9 \textwidth]{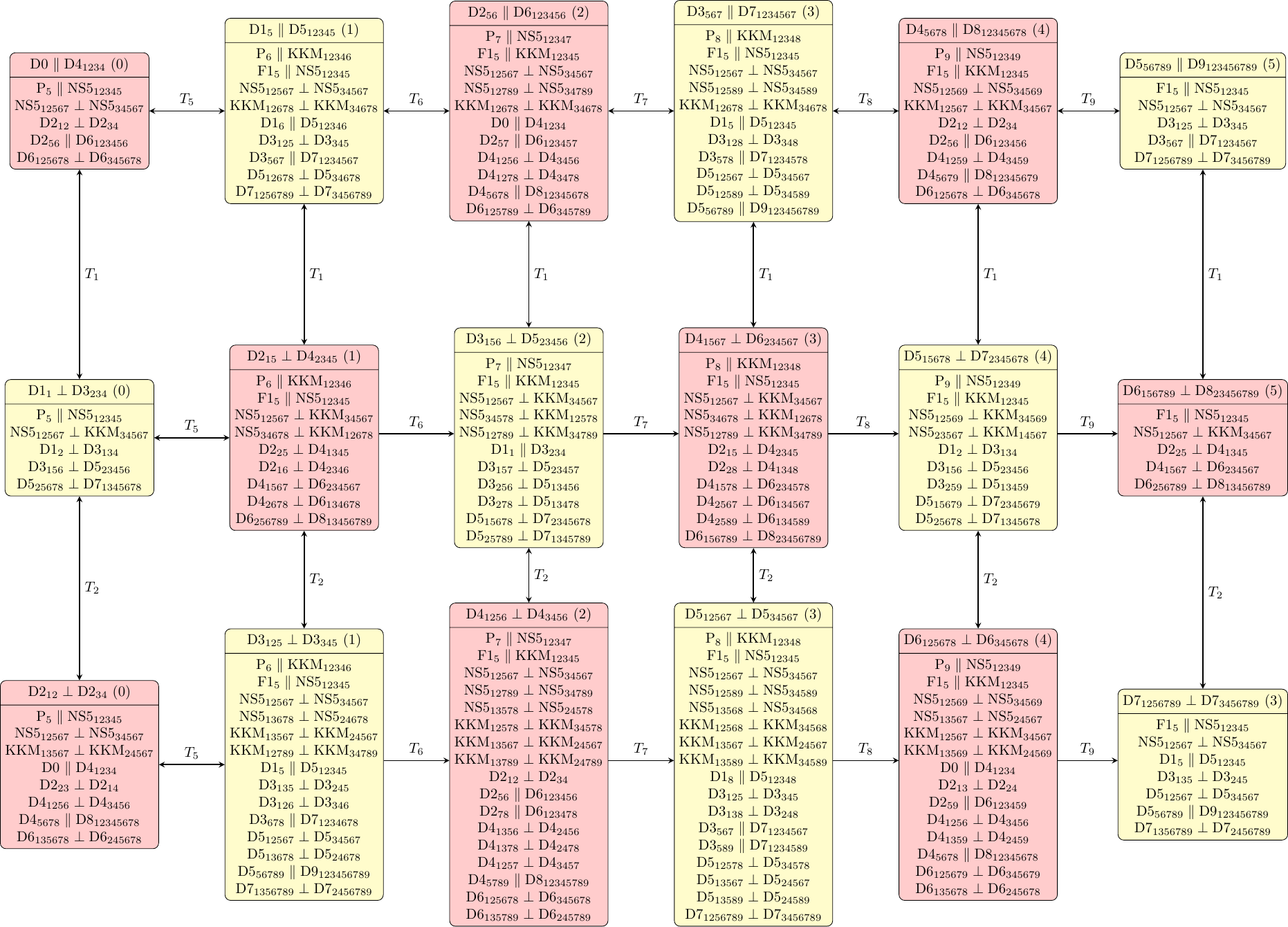}
    \caption{Possible local supersymmetry enhancements of $\D{p}$-$\D{q}$ standard configurations. For details on colouring and notation, we refer to the text, especially the segment on notation \ref{text:notation}.}
    \label{fig:Dp-Dq-Configurations}
\end{sidewaysfigure}

See Fig. \ref{fig:Dp-Dq-Configurations}.

\clearpage
\subsection{Dp - NS5 and Dq - KKM configurations}


\begin{figure}[h!]
    \centering
    \includegraphics[width = 0.7 \textwidth]{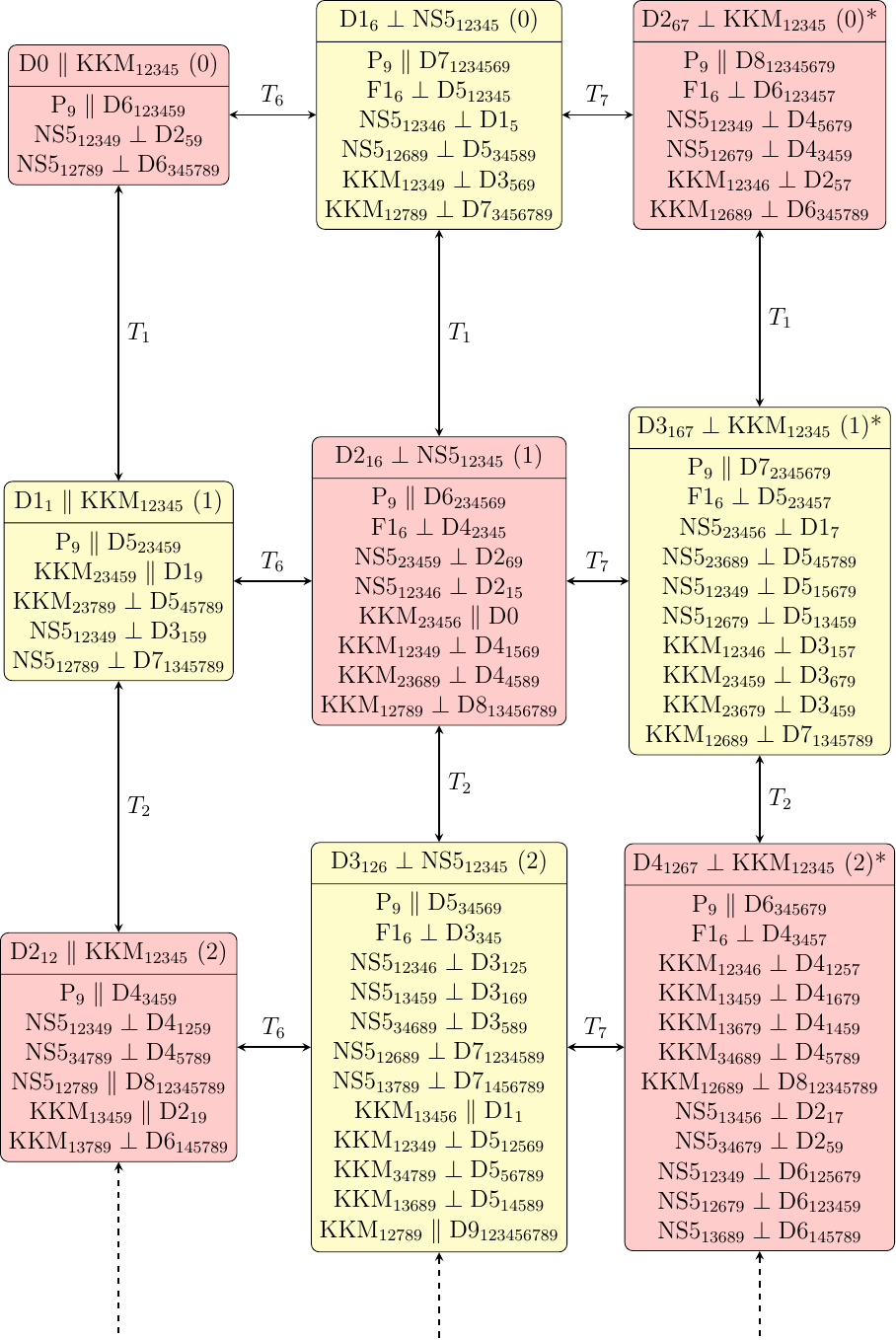}
    \caption{Possible local supersymmetry enhancements of $\D{p}$-$\NS$ and $\D{q}$-$\KKM$ standard configurations (first part). For details on colouring and notation, we refer to the text, especially the segment on notation \ref{text:notation}.}
    \label{fig:Dp-NS5-Configurations-1}
\end{figure}
\begin{figure}[h!]
    \centering
    \includegraphics[width = 0.7 \textwidth]{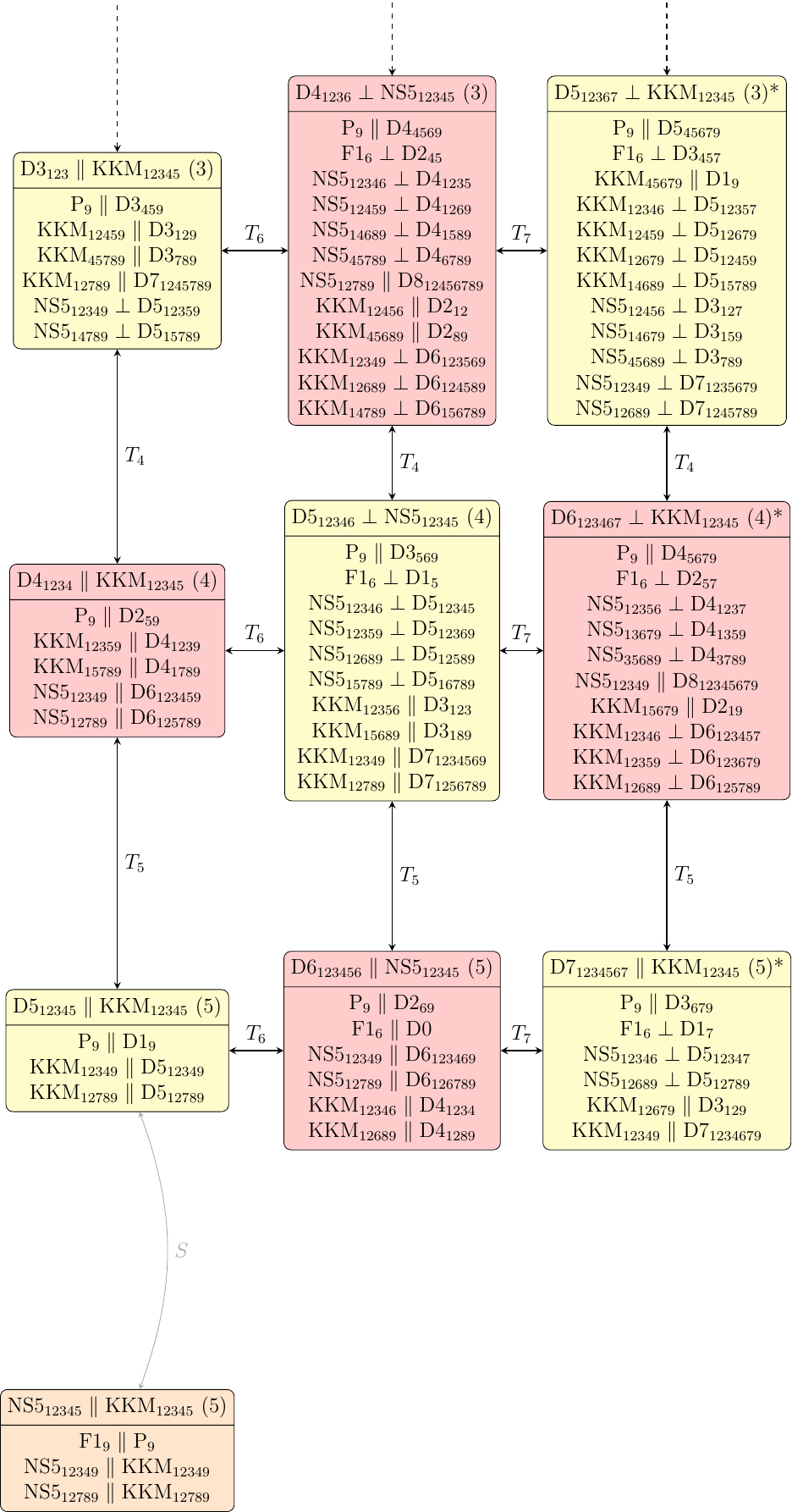}
    \caption{Possible local supersymmetry enhancements of $\D{p}$-$\NS$ and $\D{q}$-$\KKM$ standard configurations (second part). For details on colouring and notation, we refer to the text, especially the segment on notation \ref{text:notation}.}
    \label{fig:Dp-NS5-Configurations-2}
\end{figure}

\clearpage
\subsection{P/F1 - NS5/KKM configurations}

\begin{figure}[h!]
    \centering
    \includegraphics[width = 0.5 \textwidth]{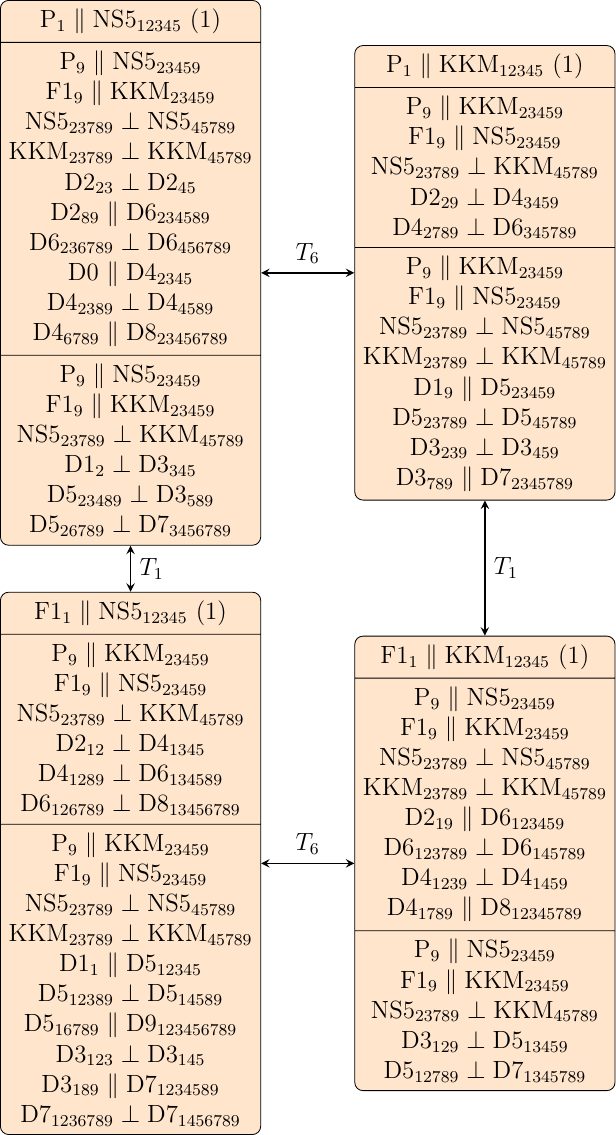}
    \caption{Possible local supersymmetry enhancements of $\W$/$\F$-$\NS$/$\KKM$ standard configurations. For details on colouring and notation, we refer to the text, especially the segment on notation \ref{text:notation}.}
    \label{fig:P-NS5-Configurations}
\end{figure}

\clearpage
\subsection{NS5/KKM - NS5/KKM configurations}


\begin{figure}[h!]
    \centering
    \includegraphics[width = 1.0 \textwidth]{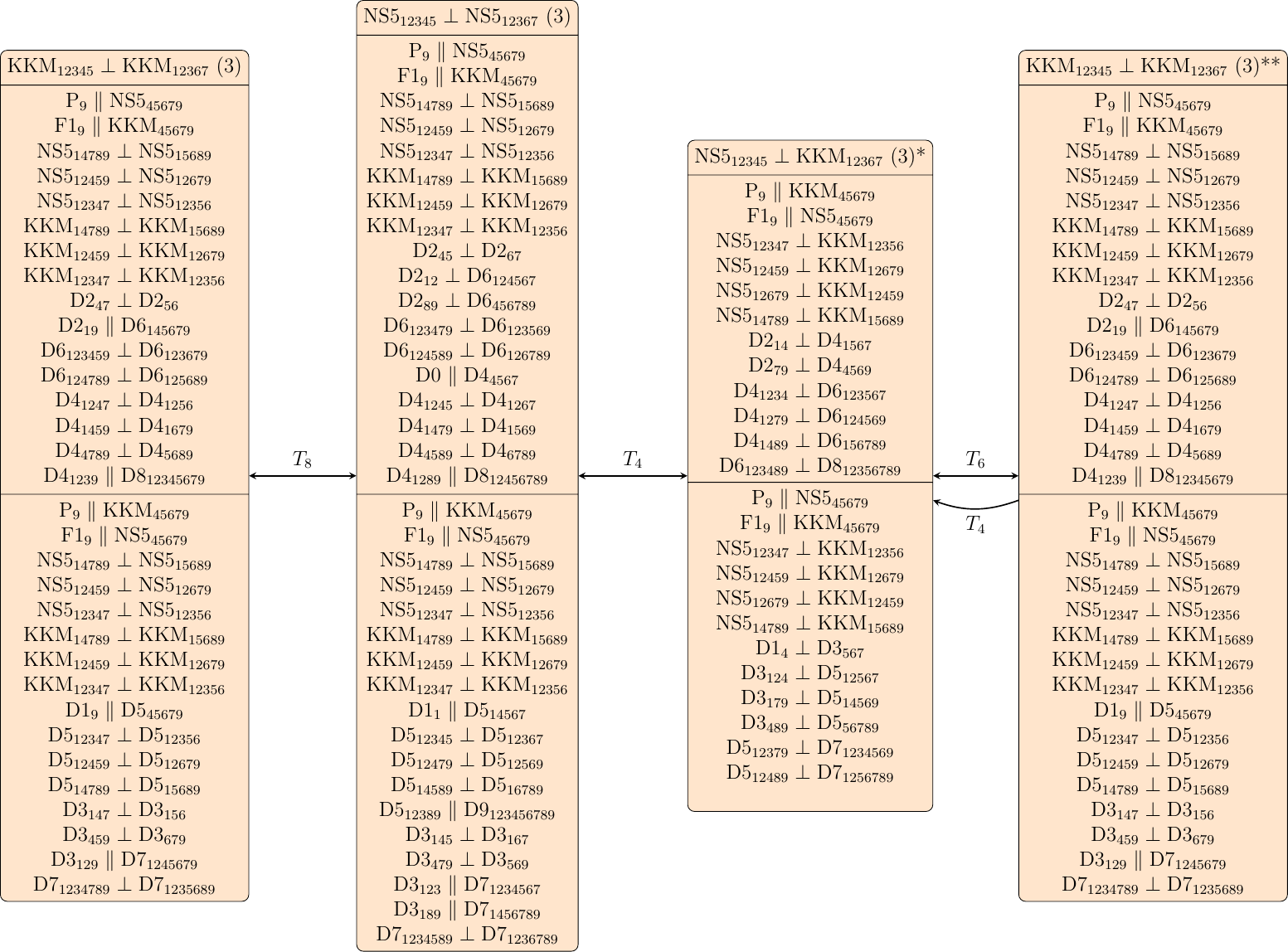}
    \caption{Possible local supersymmetry enhancements of $\NS$/$\KKM$-$\NS$/$\KKM$ standard configurations. For details on colouring and notation, we refer to the text, especially the segment on notation \ref{text:notation}.}
    \label{fig:NS5-KKM-Configurations-reordered}
\end{figure}

\clearpage
\subsection{Dp - Dq non-standard configurations}

\begin{figure}[h!]
    \centering
    \includegraphics[width = 1.0 \textwidth]{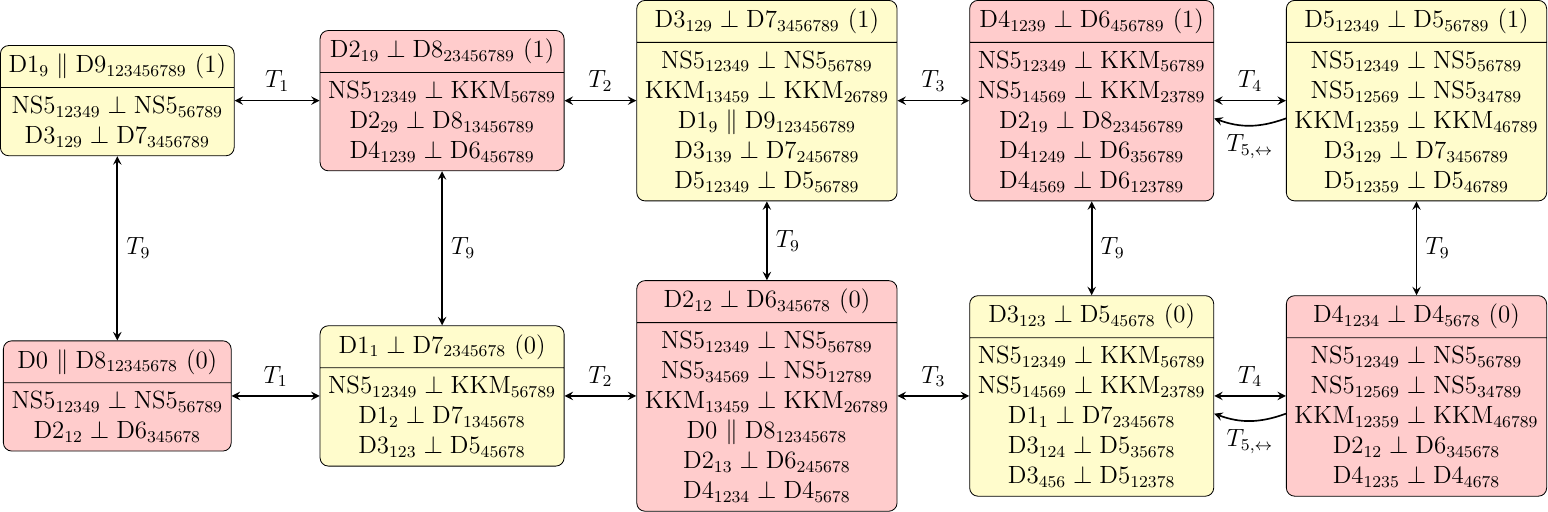}
    \caption{Possible local supersymmetry enhancements of $\D{p}$-$\D{q}$ non-standard configurations. For details on colouring and notation, we refer to the text, especially the segment on notation \ref{text:notation}.}
    \label{fig:Dp-Dq-NonStandard-Configurations}
\end{figure}

\clearpage
\subsection{NS5/KKM - NS5/KKM non-standard configurations}


\begin{figure}[h!]
    \centering
    \includegraphics[width = 1.0 \textwidth]{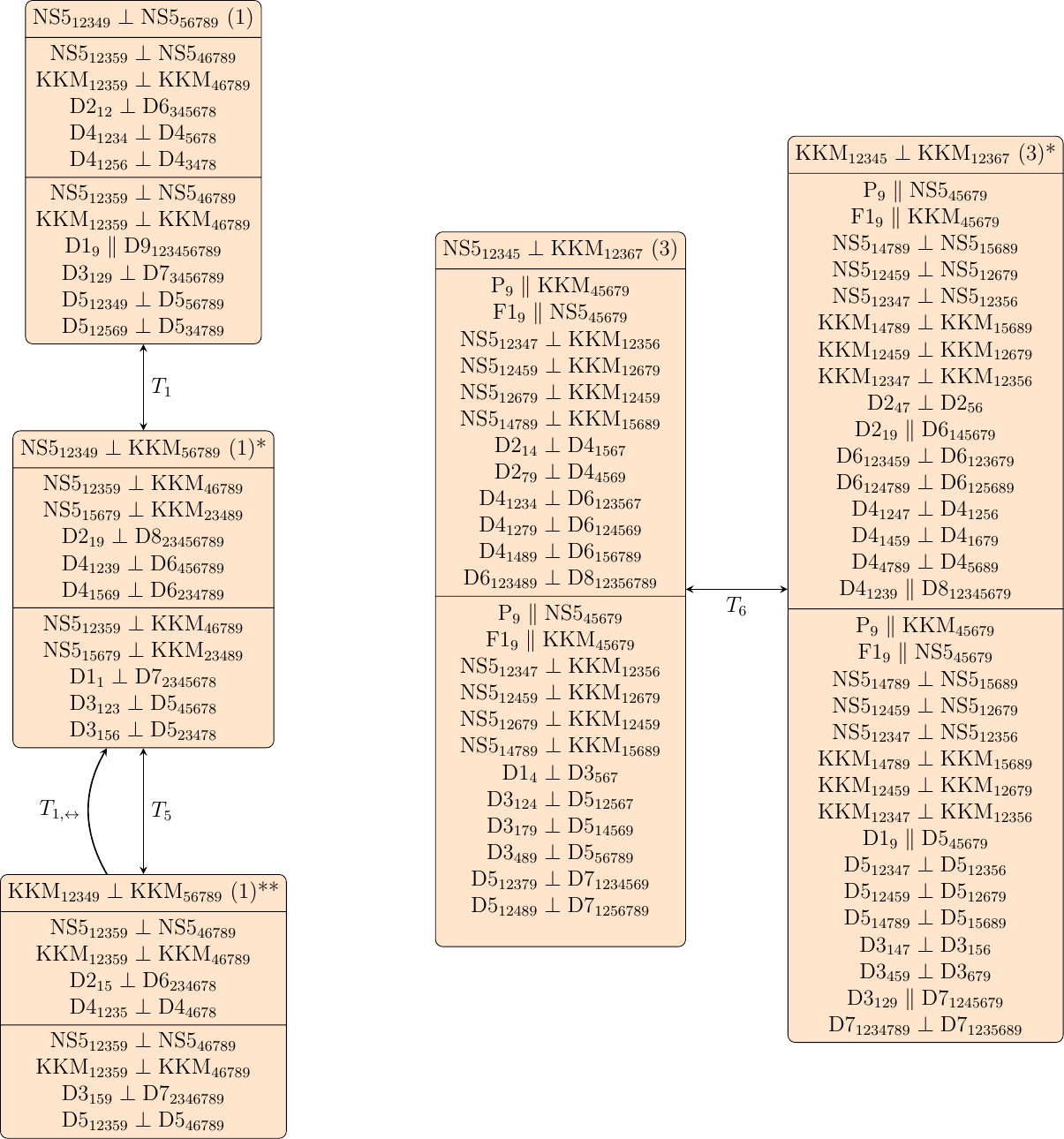}
    \caption{Possible local supersymmetry enhancements of $\NS$/$\KKM$-$\NS$/$\KKM$ non-standard configurations. For details on colouring and notation, we refer to the text, especially the segment on notation \ref{text:notation}.}
    \label{fig:NS5-KKM-NonStandard-Configurations-reordered}
\end{figure}

\clearpage
\subsection{Dp - NS5 and Dq - KKM non-standard configurations}


\begin{figure}[h!]
    \centering
    \includegraphics[width = 0.8 \textwidth]{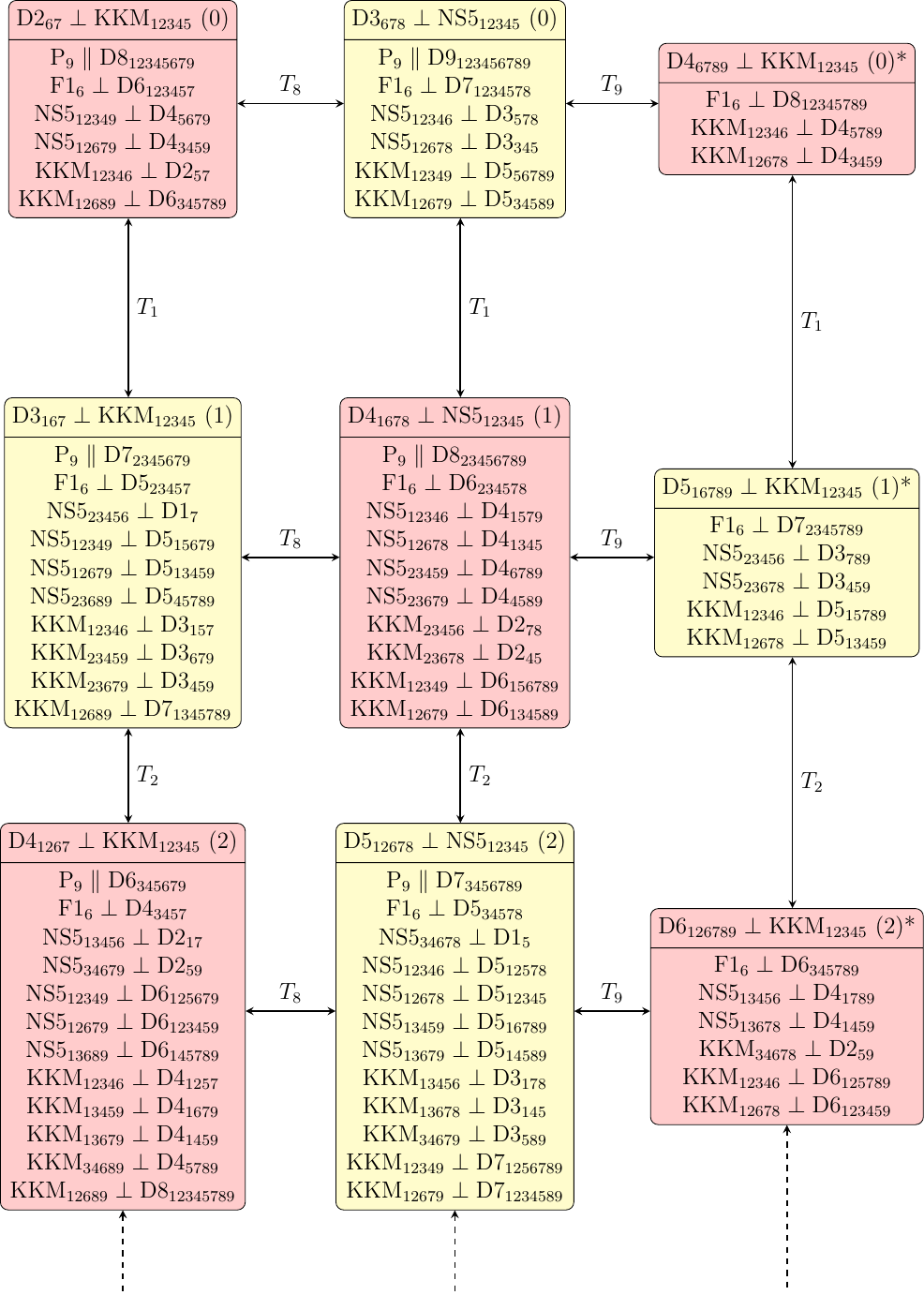}
    \caption{Possible local supersymmetry enhancements of $\D{p}$-$\NS$ and $\D{q}$-$\KKM$ non-standard configurations (first part). For details on colouring and notation, we refer to the text, especially the segment on notation \ref{text:notation}.}
    \label{fig:Dp-NS5-NonStandard-Configurations-1}
\end{figure}
\begin{figure}[h!]
    \centering
    \includegraphics[width = 0.8 \textwidth]{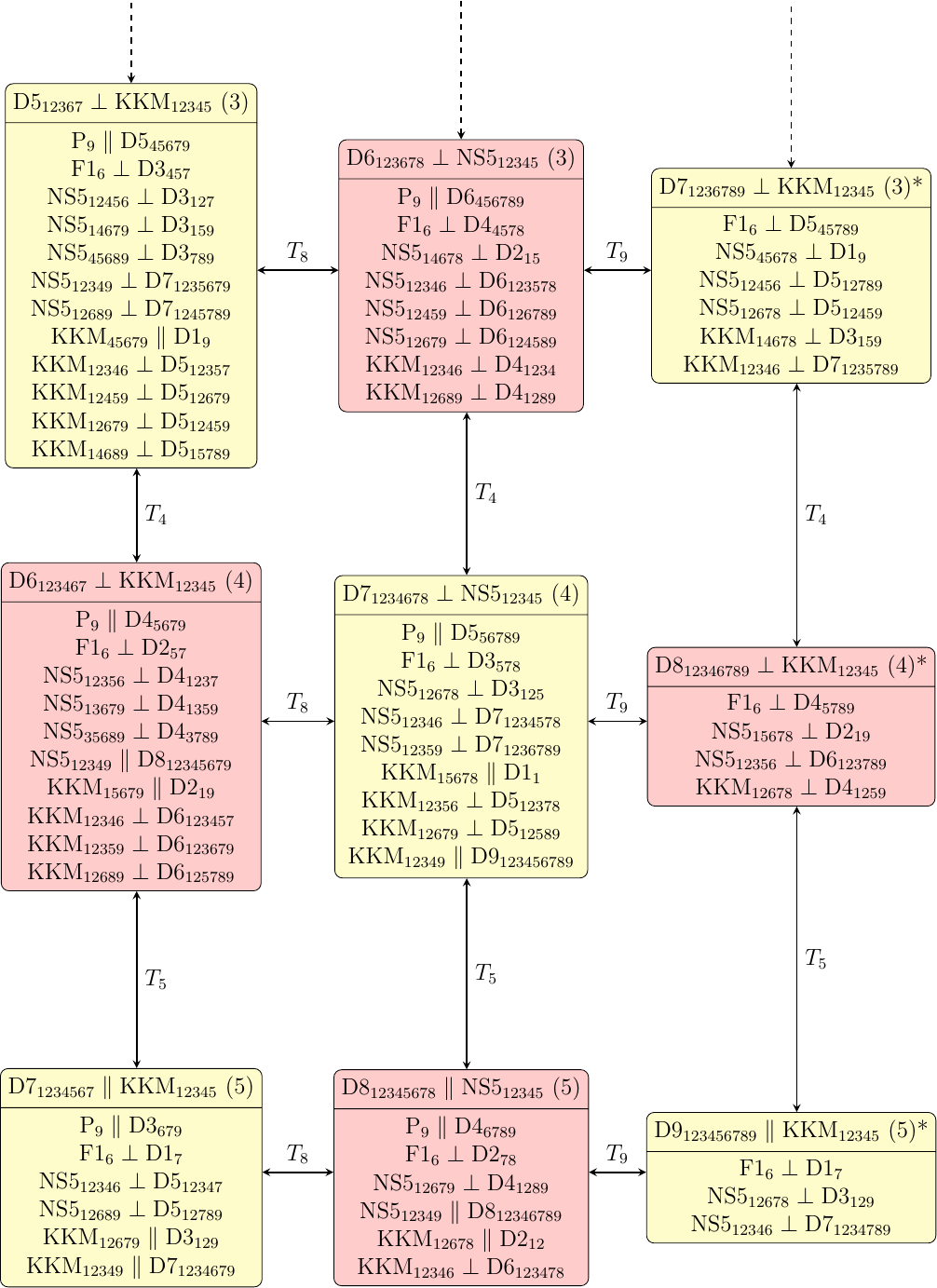}
    \caption{Possible local supersymmetry enhancements of $\D{p}$-$\NS$ and $\D{q}$-$\KKM$ non-standard configurations (second part). For details on colouring and notation, we refer to the text, especially the segment on notation \ref{text:notation}.}
    \label{fig:Dp-NS5-NonStandard-Configurations-2}
\end{figure}

\clearpage
\section{Projectors and involutions for branes}
\label{sec:projectors_and_involutions_for_branes}

Here, we give a short list of the involutions inside the supersymmetry projectors for the different types of branes. They were originally derived in \cite{Lunin:2007mj, Smith:2002wn, Kimura:2016xzd}.
The involutions for branes in M-theory are given by:
\begin{equation}
	P_{\mathrm{ P  }} = \Gamma^{01}\,, \qquad
	P_{\mathrm{ M2 }} = \Gamma^{012} \,, \qquad
	P_{\mathrm{ M5}} = \Gamma^{012345} \,, \qquad
	P_{\mathrm{ KKM(123456;7)}} = \Gamma^{0123456} \,.
\end{equation}
The involutions for the branes in Type-II string theory are given in the tables below.


\begin{table}[h]
    \centering
    \begin{tabular}{c|c c c}
        Involutions $P$ & Standard Branes & Defect Branes & Domain Walls \\ \hline
        $\Gamma^{01}$ & \W(1) & $0_4^{(1,6)}(\,,234567,1)$ &  \\
        \vspace{-.8em}\\
        $\Gamma^{01} \sigma_3$ & \F(1) & $1_4^6(1,234567)$ & \\
        \vspace{-.8em}\\
        $\Gamma^{012345}$ & \NS(12345) & $5_2^2(12345,89)$ & $5_4^3(12345,789)$ \\
        && $5_2^4(12345,6789)$ & $3_4^{(2,3)}(123,789,45)$ \\
        &&& $1_4^{(4,3)}(1,789,2345)$ \\
        \vspace{-0.8em}\\
        $\Gamma^{012345} \sigma_3$ & \KKM(12345) & $5_2^3(12345,789)$ & $4_4^{(1,3)}(1234,789,5)$ \\
        &&& $2_4^{(3,3)}(12,789,345)$ \\
        &&& $0_4^{(5,3)}(\,,789,12345)$ 
    \end{tabular}
    \caption{Involutions inside the supersymmetry projectors for different standard branes, exotic branes and domain walls in Type-IIA string theory.}
    \label{tab:ExoticBraneInvolutionsForSolitonicBranesIIA}
\end{table}

\begin{table}[h]
    \centering
    \begin{tabular}{c|c c c}
        Involutions $P$ & Standard Branes & Defect Branes & Domain Walls \\ \hline
        $\Gamma^{01}$ & \W(1) & $0_4^{(1,6)}(\,,234567,1)$ &  \\
        \vspace{-.8em}\\
        $\Gamma^{01} \sigma_3$ & \F(1) & $1_4^6(1,234567)$ & \\
        \vspace{-.8em}\\
        $\Gamma^{012345}$ & \KKM(12345) & $5_2^3(12345,789)$ & $5_4^3(12345,789)$ \\
        &&& $3_4^{(2,3)}(123,789,45)$ \\
        &&& $1_4^{(4,3)}(1,789,2345)$ \\
        \vspace{-0.8em}\\
        $\Gamma^{012345} \sigma_3$ & \NS(12345) & $5_2^2(12345,89)$ & $4_4^{(1,3)}(1234,789,5)$ \\
        && $5_2^4(12345,6789)$ & $2_4^{(3,3)}(12,789,345)$ \\
        &&& $0_4^{(5,3)}(\,,789,12345)$ 
    \end{tabular}
    \caption{Involutions inside the supersymmetry projectors for different standard branes, exotic branes and domain walls in Type-IIB string theory.}
    \label{tab:ExoticBraneInvolutionsForSolitonicBranesIIB}
\end{table}

\begin{table}[h]
    \centering
    \begin{tabular}{c|c c c}
        Involutions $P$ & Standard Branes & Defect Branes & Domain Walls \\\hline
        $\Gamma^{0} i \sigma_2$ & $\D{0}$ & $0_3^7(\,,1234567)$ &  \\
        $\Gamma^{01} \sigma_1$ & $\D{1}(1)$ & $1_3^6(1,234567)$ & $0_3^{(1,7)}(\,,1234567,8)$ \\
        $\Gamma^{012} \sigma_1$ & $\D{2}(12)$ & $2_3^5(12,34567)$ & $1_3^{(1,6)}(1,234567,8)$ \\
        $\Gamma^{0123} i \sigma_2$ & $\D{3}(123)$ & $3_3^4(123,4567)$ & $2_3^{(1,5)}(12,34567,8)$ \\
        $\Gamma^{01234} i \sigma_2$ & $\D{4}(1234)$ & $4_3^3(1234,567)$ & $3_3^{(1,4)}(123,4567,8)$ \\
        $\Gamma^{012345} \sigma_1$ & $\D{5}(12345)$ & $5_3^2(12345,67)$ & $4_3^{(1,3)}(1234,567,8)$ \\
        $\Gamma^{0123456} \sigma_1$ & $\D{6}(123456)$ & $6_3^1(123456,7)$ & $5_3^{(1,2)}(12345,67,8)$ \\
        $\Gamma^{01234567} i \sigma_2$ & $\D{7}(1234567)$ & $7_3(1234567)$ & $6_3^{(1,1)}(123456,7,8)$ \\
        $\Gamma^{012345678} i \sigma_2$ & $\D{8}(12345678)$ &  & $7_3^{(1,0)}(1234567,,8)$ \\
        $\Gamma^{0123456789} \sigma_1$ & $\D{9}(123456789)$ &  &  \\
    \end{tabular}
    \caption{Involutions inside the supersymmetry projectors for different D-branes, defect branes and domain walls in Type-II string theory.}
    \label{tab:ExoticBraneInvolutionsForDBranes}
\end{table}

\clearpage

\bibliographystyle{JHEP}

\bibliography{mainLSE}

\end{document}